\documentclass[aps,prl,superscriptaddress,float fix,10pt,twocolumn]{revtex4-1}
\usepackage{graphicx,graphics}
\usepackage{dcolumn}
\usepackage{amsmath,amssymb,amsfonts}
\usepackage{latexsym,verbatim}
\usepackage{physics}
\usepackage{bm}
\usepackage{color}
\usepackage{cancel}
\usepackage[normalem]{ulem}
\usepackage{multirow}
\usepackage[abs]{overpic}
\usepackage[breaklinks=true,color links,citecolor=blue,linkcolor=blue,urlcolor=blue]{hyperref}
\usepackage{array}
\usepackage{nicefrac}
\usepackage{dsfont}
\usepackage{float}
\usepackage{tikz}
\usepackage{overpic}
\usepackage{caption}
\usepackage{subcaption}
\usepackage{ragged2e}

\begin{document}
\title{Bogoliubov excitations of a Bose-Einstein condensate in the presence of the non-Hermitian skin effect}
\author{Carlo Panu}
\affiliation{Institute of Condensed Matter Theory and Solid State Optics, Friedrich-Schiller-Universit\"at Jena,
Max-Wien-Platz 1, 07743 Jena,~Germany}
\affiliation{Department of Quantum Science and Technology, Research School of Physics, The Australian National University, Canberra ACT 2601, Australia}



\author{Ulf Peschel}
\affiliation{Institute of Condensed Matter Theory and Solid State Optics, Friedrich-Schiller-Universit\"at Jena,
Max-Wien-Platz 1, 07743 Jena,~Germany}

\author{Oleg A. Egorov}
\affiliation{Institute of Condensed Matter Theory and Solid State Optics, Friedrich-Schiller-Universit\"at Jena,
Max-Wien-Platz 1, 07743 Jena,~Germany}

\author{Elena A. Ostrovskaya}
\affiliation{Department of Quantum Science and Technology, Research School of Physics, The Australian National University, Canberra ACT 2601, Australia}

\author{Eliezer Estrecho}
\affiliation{Department of Quantum Science and Technology, Research School of Physics, The Australian National University, Canberra ACT 2601, Australia}
\affiliation{School of Science, The University of New South Wales, Canberra ACT 2601, Australia}

\begin{abstract}
The non-Hermitian skin effect is one of the most remarkable topological features of non-conservative physics. It causes a macroscopic accumulation of states at the boundaries of a confined system, opening new ways for controlling and directing waves in non-Hermitian settings. This effect has recently been observed in experiments with Bose-Einstein condensates of exciton polaritons and ultracold bosonic quantum gases. The underlying physics is captured by a mean-field Hamiltonian with a trapping potential and an imaginary gauge field. While prior work has examined the effect of the imaginary gauge field term on the condensate itself, its effect on the collective excitations on top of the condensate remained unexplored. Here, we develop a new theory for the Bogoliubov excitations in a non-Hermitian framework and systematically study their spectrum across different interaction regimes. By analysing the position expectation values, we find that while the repulsive interactions oppose the localisation of the condensate due to the skin effect, the excitations remain strongly affected by the imaginary gauge field. As a consequence, in the presence of harmonic confinement, we predict a strong shift in the dipole oscillation frequency, which can be observed in experiments. This is in contrast to Hermitian systems, where this frequency is fixed by the trap dimensions. Furthermore, we develop a consistent second-quantised field theory of the collective excitations in this non-Hermitian system. Strikingly, we find that a second-quantised treatment removes the  shift of the dipole oscillation frequency entirely. This intriguing discrepancy between the two theoretical approaches to the treatment of elementary excitations in a non-Hermitian system can be resolved in experiments.
\end{abstract}

\maketitle

{\it Introduction.---} The skin effect is one of the main features of the non-Hermitian tight-binding model introduced by N. Hatano and D. R. Nelson \cite{HN1996}. Their work aimed to model a bosonic system with a random potential describing defects in superconductors. The starting point was the typical continuous Schr\"odinger equation with the addition of an imaginary vector potential within the minimal coupling description \cite{Nelson1990}. In order to perform numerical calculations, the model has been translated to a tight-binding lattice model with asymmetric hopping terms. More recently, this discrete version gained popularity in the context of non-Hermitian lattice models because it is the simplest model that can exhibit macroscopic accumulation of eigenstates near the boundary \cite{Kunst2018}, \textit{i.e.}, the non-Hermitian skin effect (NHSE). The topological classification has been extended to systems with these features \cite{Gong2018, Kawabata2019}, and the novel non-Hermitian topology has been the subject of interest in the past years \cite{Okuma2023}. Motivated by the progress of theory, the analysis of imaginary fields in the continuum counterparts of the discrete models has been revisited, predicting localisation effects analogous to the skin effect \cite{Longhi2021, Yuce2023}. These studies were motivated by the goal of finding a more general description of a nearly free particle, in the limit when the tight-binding approximation breaks down.

Recent experimental studies made use of continuous models with an imaginary gauge potential to describe physical systems such as bosonic gases of exciton-polaritons \cite{Robin2025,Liang2026} and cold atoms \cite{Tao2026,liang2022,zhao2025two}. These physical systems enable experimental observation of the eigenstates, which makes them an attractive testbed for theory. In both systems, the imaginary gauge potential can be introduced by tunable, asymmetric gain and/or loss. However, due to the interaction-driven formation of the macroscopically occupied single-particle quantum state, the Bose-Einstein condensate, the two systems are intrinsically nonlinear, which raises the question of the interplay between the NHSE-induced localisation and mean-field interactions. Moreover, recent experimental studies of electric circuit metamaterials \cite{Lo2025} suggest that the interactions-induced nonlinearity is responsible for inducing strong NHSE in the linear excitation modes on top of the system's ground state, the phenomenon that could also occur in the behaviour of collective excitations of bosonic condensate.

Here, we study the interplay between the NHSE-induced localisation of a trapped BEC and mean-field interactions in a simplified model. We first look at the nonlinear stationary states of a Gross-Pitaevskii \cite{pitaevskii2016} equation with a complex gauge potential and different confining potentials. Subsequently, we study the elementary (Bogoliubov) excitations of the condensate by utilising the Bogoliubov-de Gennes (BdG) approach that, at the same time, allows us to test the stability of the stationary states and characterise the unstable ones \cite{supplemental}. Insights about the NHSE are then obtained by tracking the position expectation value $\langle x \rangle$ of both the nonlinear states and BdG excitations while varying the complex gauge and the nonlinear coupling strengths. Our results show that repulsive interaction ``screens'' the NHSE of the nonlinear states while attractive interaction enhances it. However, even in the Thomas-Fermi regime where the repulsive nonlinearity fully screens the NHSE, the localisation persists in the Bogoliubov-de Gennes modes and leads to a modification of their excitation frequencies. Interestingly, the latter phenomenon is not present when the excitations are treated using a quantum field approach developed for this non-Hermitian scenario. Finally, we discuss the physical implications of the results for both equilibrium (cold atoms) and non-equilibrium (polaritons) condensates.

{{\it Model.---}} Similarly to Refs. \cite{Robin2025, Longhi2021}, we start from a single-band, one-dimensional single-particle non-Hermitian Hamiltonian with minimal coupling to the purely imaginary gauge field $i\beta$, defined as follows (see \cite{supplemental} for derivation from a more familiar quantum mechanical version): 
\begin{equation}
      \hat{H}_{\mathrm{nH}} \equiv -\frac{1}{2}\big(\partial_x - \beta\big)^2 + V_{\mathrm{trap}}(x) \, ,
      \label{eq:ham_nh}
\end{equation}
where $V_{\mathrm{trap}}(x)$ is a real confining potential. Here, the Hamiltonian is rescaled to dimensionless units \footnote{If we consider for example an harmonic trapping potential, one has $x = x' / l$, with $ l = \sqrt{\hbar/m\omega}$ and then divides $H_{\mathrm{nH}}$ by $\hbar \omega$.}.
The Hamiltonian is related to a Hermitian operator by the similarity transformation $S = \mathrm{exp}(\beta x)$,
\begin{equation}
    \hat{H} = S \hat{H}_{\mathrm{nH}} S^{-1} = -\frac{1}{2} \partial^2_x +  V_{\mathrm{trap}}(x) \, ,
\label{eq:gauge_transf}
\end{equation}
resulting in modified localisation properties of the eigenfunctions 
\begin{equation}
    \Phi_{\mathrm{nH}} = e^{+\beta x} \Phi,
\end{equation}
but leaving the eigenvalues unchanged.
This transformation works for all $\beta$ as long as the confining potential compensates the exponential growth and ensures a faster spatial decay of the wavefunction. The exponential localisation of a macroscopic number of eigenstates (all eigenstates in this case) is one of the defining features of the NHSE skin effect, which is accompanied by a point-gap topology of the eigenvalues in complex space~\cite{Robin2025, Longhi2021}. In the following, we explore how interactions modify the localisation of both the eigenstates and their elementary excitations.

{{\it Mean-field theory.---}} In the presence of interactions, the Gross-Pitaevskii \cite{pitaevskii2016} equation (GPE) describing the mean-field $c$-number wavefunction is:
\begin{equation}
    i \partial_t \Phi = \big[\hat{H}_{\mathrm{nH}} + g |\Phi|^2 \big] \Phi, 
    \label{eq:gross_pita}
\end{equation}
where the linear Hamiltonian is defined by Eq.~(\ref{eq:ham_nh}). We seek stationary solutions of the form $\Phi_s(x, t)~=~\Phi_s(x)~\mathrm{exp}(-i\mu t)$, leading to a nonlinear eigenvalue problem. Linear stability of the eigenstates $\Phi_s$ is determined by introducing small, generally complex-valued, fluctuations $\delta\Phi$:
\begin{equation}
    \Phi(x, t) =\Big[\Phi_s(x) + \delta\Phi(x, t)\Big] e^{-i\mu t} \, .
    \label{eq:stationary}
\end{equation} 
Substituting into Eq. (\ref{eq:gross_pita}) and collecting terms to the first order in $\delta \Phi(x, t)$ and $\delta \Phi^*$, one obtains the linearised problem: 
\begin{equation}
    i\partial_t 
    \begin{pmatrix}
        \delta \Phi \\
        \delta \Phi^*
    \end{pmatrix} =  \hat{\mathcal{L}} \begin{pmatrix}
        \delta \Phi \\
        \delta \Phi^*
    \end{pmatrix} \, ,
    \label{eq:bdg_problem}
\end{equation}
where the BdG operator is defined as:
\begin{equation}
    \hat{\mathcal{L}} = \begin{pmatrix}
        \hat{H}_{\mathrm{nH}} -\mu + 2g |\Phi_s|^2 & -g \Phi_s^2 \\
        g(\Phi_s^*)^2 & -\big[\hat{H}_{\mathrm{nH}} -\mu + 2g |\Phi_s|^2\big]^* 
    \end{pmatrix} .
    \label{eq:bdg_operator_mt}
\end{equation}
In this work, we focus on the spectral features of $\mathcal{L}$ in the presence of the imaginary gauge field. Following the standard Bogoliubov prescription \cite{castin2002bose}, we decompose the linear fluctuation in terms of stationary solutions 
\begin{equation}
    \delta \Phi(x, t) = \sum_i \Big[ u_i(x) e^{-i\epsilon_i t} - v^*_i(x) e^{+i\epsilon_i t} \Big].
    \label{eq:bdg_prescription}
\end{equation}
Matching the exponentials Eq. \ref{eq:bdg_problem} results in an eigenvalue problem.
In the general non-Hermitian framework, we can define \cite{giacobino2025} the right ($r$) and left ($l$) eigenvectors of $\mathcal{L}$ as follows \footnote{Throughout, we omit the subscript $r$ on right eigenvectors for brevity, writing $\mathbf{w}$, $u$ and $v$ in place of $\mathbf{w_r}$, $u_r$ and $v_r$}: 
\begin{equation}
    \hat{\mathcal{L}} \mathbf{w}^{i} = \epsilon_i \mathbf{w}^{i} \quad \mathrm{and} \quad \hat{\mathcal{L}}^{\ddagger} \mathbf{w}^{i}_l = \epsilon_i \mathbf{w}^{i}_l \ ,
\end{equation}
where $\hat{\mathcal{L}}^{\ddagger}~=~\sigma_3 \hat{\mathcal{L}}^{\dagger}\sigma_3$, the Pauli matrix $\sigma_3~=~\mathrm{diag}(1, -1)$ and we defined the vector $\textbf{w}= (u, \, v)^T$. The left and right eigenvectors make a complete basis set of quasiparticles and obey the biorthogonal normalisation conditions:
\begin{equation}
    \int dx \, \mathbf{w}^{i}_l(x) \, \sigma_3 \, \mathbf{w}^{j}(x) = \delta_{ij} \, .
    \label{eq:biorthogonality}
\end{equation}
In the case under investigation, due to confinement, we have a pure non-Hermitian gauge term, and the spectral symmetries and properties of $\hat{\mathcal{L}}$ are the same as those in the Hermitian limit $\beta \rightarrow 0$ ($\hat{\mathcal{L}}_0 $) \cite{castin2002bose}. The gauge dressing can be combined with the Nambu space \cite{Mbeng2024}, and we obtain a new set of transformations with $\eta_i = \sigma_i \mathcal{S}^{-2}$, with $\mathcal{S} = \mathrm{diag}(S, S)$. In addition, we have $\sigma_3 \hat{\mathcal{L}}_0 \sigma_3 = \hat{\mathcal{L}}_0^{\dagger}$, equivalent to
\begin{equation}
     \hat{\mathcal{L}}^{\dagger} = \mathcal{S}^{-1} \hat{\mathcal{L}}^{\dagger}_0 \mathcal{S} =  \mathcal{S}^{-1} \sigma_3 \hat{\mathcal{L}}_0 \sigma_3 \mathcal{S} = \mathcal{S}^{-2}\sigma_3 \hat{\mathcal{L}} \sigma_3 \mathcal{S}^2,
\end{equation}
where the operator $\mathcal{S}=\mathcal{S}^{\dagger}$ commutes with $\sigma_3$. As discussed above, the transformation does not act on the eigenvalues, so $\hat{\mathcal{L}}$ and $\hat{\mathcal{L}}^{\dagger}$ have the same spectrum, meaning that if $\epsilon$ is an eigenvalue, then $\epsilon^*$ is as well. On the other hand, if we name the right eigenvector components, following the usual convention, $\mathbf{w}=(u, v)^T = \mathrm{exp}(+\beta x)(u_0, v_0)^T$, the eigenvectors of $\hat{\mathcal{L}}^{\dagger}$ will exhibit an exponential localisation on the opposite side: $\mathbf{w}_{l} = \mathrm{exp}(-\beta x)(u^*_0, -v^*_0)^T$, as one could expect from the left eigenvectors in the Hatano-Nelson model. 
The additional symmetry $\sigma_1 \hat{\mathcal{L}} \sigma_1 = - \hat{\mathcal{L}}^*$, where $\sigma_1$ is the first Pauli matrix, implies that $\hat{\mathcal{L}}$ also has an eigenvector $(v^*, u^*)^T$ with an eigenvalue $-\epsilon^*$, analogous to the particle-hole symmetry for the bosonic case. The stability condition for the stationary solution $\Phi_s(x,t)$ then becomes $\mathrm{Im}\{\epsilon_i\}=0$ $\forall \ i$. 

To express the excitations of Eq. (\ref{eq:bdg_prescription}) in terms of experimentally accessible observables, we employ the so-called Madelung transformation to hydrodynamic form $\Phi(x, t) = \sqrt{n(x, t)} \, \mathrm{exp}\{i\vartheta(x, t)\}$ \cite{pitaevskii2016, Stringari1996}, which allows us to derive the density and phase fluctuation profiles directly from the quasiparticle amplitudes. To this end, we write a solution of the stationary GPE, introduce the fluctuations of density, $\delta n$ and phase, $\delta \vartheta$, and expand up to first order in these fluctuations:
\begin{equation}
    \delta \Phi(x, t) =  \bigg[\frac{\delta \, n(x, t)}{2 n_s(x)} + i \delta \vartheta (x, t)\bigg] \sqrt{n_s(x)} e^{i\vartheta_s(x)}\,.
\end{equation}
Making a decomposition into stationary waves \footnote{Notice that, while $\delta n(x)$ and $\delta \vartheta(x)$ and be complex, $\delta n(x, t)$ and $\delta \vartheta(x, t)$ are real.}
\begin{equation}
  \delta n(x, t) = \sum_i \Big[ \delta n_i(x) e^{-i\epsilon_i t} + \delta n^*_i(x) e^{+i\epsilon_i t} \Big] 
\end{equation}
and similarly for $\vartheta(x, t)$, we can then compare with Eqs. (\ref{eq:bdg_prescription}) and (\ref{eq:stationary}), obtaining the following relations
\begin{equation}
\begin{aligned}
       \delta n_i (x) =& \sqrt{n_s(x)} \bigg[u_i(x)e^{-i\vartheta_s(x)}  - v_i(x)e^{i\vartheta_s(x)}\bigg], \\
       \delta \vartheta_i (x) =& \frac{1}{2i\sqrt{n_s(x)}} \bigg[u_i(x)e^{-i\vartheta_s(x)}  + v_i(x)e^{i\theta_s(x)}\bigg] \, .
\end{aligned}
\label{eq:density_phase}
\end{equation}
An alternative and often more transparent approach to the collective excitation spectrum is offered by the Thomas-Fermi (TF) approximation combined with a hydrodynamic reformulation of the dynamics \cite{pitaevskii2016}. The TF regime is reached when $g\gg 1$, allowing one to neglect the kinetic energy term in the GPE and leading to the following expression for the ground state density of the nonlinear problem:
\begin{equation}
    n_{\mathrm{TF}}(x) = \frac{\tilde{\mu} - V(x)}{g} \theta(\tilde{\mu}-V(x)) \, ,
\end{equation}
where $\tilde{\mu} = \mu + \beta^2/2$ and $\theta(x)$ is the Heaviside function.  
In this regime, rather than the diagonalizing the BdG operator, one can solve the following modified hydrodynamic equations for a superfluid:
\begin{equation}
    (\partial_x - 2\beta) \big[g n_{\mathrm{TF}}(x) \, \delta n(x)\big] = -\epsilon^2 \delta n(x)
    \label{eq:hydrodynamic_TF}
\end{equation}
obtained starting from the Madelung form and combining two coupled equations for the density and the superfluid velocity field $\mathrm{v} = \,\partial_x\vartheta$ \footnote{See the supplementary material.}. 
The hydrodynamic approach therefore provides both an analytic benchmark for the BdG spectrum and a physically intuitive picture of the excitations.

{\it {Quantum theory.---}} Within the mean-field approximation, the condensate is a classical field obeying the Gross–Pitaevskii equation, and linearisation yields the Bogoliubov–de Gennes spectrum of collective modes. This spectrum, however, is a property of the linearised operator alone and does not require a quantum interpretation of its amplitudes. Reformulating the fluctuation field in terms of a quasiparticle expansion adds the effects not captured by the mean-field description: the interacting ground state acquires a finite quantum depletion \cite{Lopes2017,Pieczarka2020}, the modes inherit bosonic statistics and correlations properties that can be directly probed in experiments~\cite{Pieczarka2020,Utsunomiya2008,Claude2023,Pieczarka2015,Bieganska2021}. In the non-Hermitian setting, further subtleties arise and a consistent quantum treatment of the non-Hermitian BdG problem is necessary. 

Here, we consider the theory of a dilute gas \cite{Fetter1972, Lewenstein1996} in the presence of an imaginary gauge field, following the treatment in Refs. \cite{Yamamoto2019, Kawabata2020}. In particular, we study the second-quantized Hamiltonian operator $\hat{\mathcal{K}} = \hat{\mathcal{H}} - \mu \hat{\mathcal{N}}$ with $\hat{\mathcal{N}} = \int dx \, \hat{\overline{\Phi}}(x, t)\hat{\Phi}(x, t)$, where $\hat{\overline{\Phi}}$ and $\hat{{\Phi}}$ are matter degrees of freedom fulfilling the bosonic commutation relations, and 
\begin{equation}
        \hat{\mathcal{H}} = \int dx \hat{\overline{\Phi}} \, \hat{H}_{\mathrm{nH}} \, \hat{\Phi} \, + \,\frac{g}{2}\int dx \, \hat{\overline{\Phi}} \,  \hat{\overline{\Phi}} \, \hat{\Phi} \, \hat{\Phi} \, ,
\end{equation}
with $\hat{H}_{\mathrm{nH}}$ defined in Eq. (\ref{eq:ham_nh}). We seek, as before, solutions in the form of perturbed stationary states $\hat{\Phi}(x)~=~\Phi_s(x)~+~\delta\hat{\Phi}(x)$  \footnote{We consider the time-independent case, equivalent to making the following tranformation \cite{Mbeng2024}: $(u, v)^T \rightarrow (u(t), v(t))^T=\mathrm{exp}\{-i\epsilon t\}(u, v)^T$ and $(\overline{u}, \overline{v})^T \rightarrow (\overline{u}(t), \overline{v}(t))^T=\mathrm{exp}\{+i\epsilon t\}(\overline{u}, \overline{v})^T$.} and, similarly, $\hat{\overline{\Phi}}(x)~=~\overline{\Phi}_s(x)~+~\delta\hat{\overline{\Phi}}(x)$. The perturbations on top of $c$-number wave functions $\Phi_s$ and $\overline{\Phi}_s$, are solutions of the following set of GPEs (see \cite{supplemental}), obtained by imposing the saddle point condition for the action \cite{supplemental, Yamamoto2019, Kamenev2023}:
\begin{equation}
    \begin{aligned}
    \big(&\hat{H}_{\mathrm{nH}} + g\rho_s - \mu \big) \Phi_s = 0 \\
    \big(&\hat{H}_{\mathrm{nH}}^T + g \rho_s - \mu \big) \overline{\Phi}_s = 0 \, , 
    \end{aligned}
    \label{eq:gross_pita_quantum}
\end{equation}
where $\rho_s(x) = \overline{\Phi}_s(x) \Phi_s(x)$ obeys the normalization condition $\int \rho_s(x) dx = 1$ \cite{Yamamoto2019}. The transpose operation is formally defined as
\begin{equation}
    \int dx \psi \hat{H} \phi = \int dx \phi \hat{H}^T \psi \, 
\end{equation}
and, contrary to the complex conjugate, the formal transpose flips the sign of the imaginary gauge.

This result was previously obtained in the linear tight-binding framework \cite{Kawabata2019, Yokomizo2021} and has an important consequence, because the imaginary gauge is now ``removable''in the GPEs similarly to what happens in the linear case, see Eq. (\ref{eq:gauge_transf}). The two stationary solutions can be written in terms of the ones in the Hermitian limit $\Phi_0$ as $\Phi_s = S \Phi_0 $ and $\overline{\Phi}_s = S^{-1} \Phi_0$ with $S = \mathrm{exp}\{\beta x\}$; as a consequence the quantities $\hat{\mathcal{N}}$ and $\rho_s$ do not depend on $\beta$ and are not affected by the NHSE. Hence, $\rho_s$ differs from the classical density $n$ introduced above.

We can now re-write the second-quantised Hamiltonian, excluding the Goldstone mode \cite{Lewenstein1996}, in the form:
\begin{equation}
    \hat{\mathcal{K}} =  E_s ( \Phi_s, \overline{\Phi}_s) \, + \, \sum_i \epsilon_i \hat{\overline{\gamma}}_i \hat{\gamma}_i
\end{equation}
where $\hat{\gamma}$ and $\hat{\overline{\gamma}}$ are quasi-particle annihilation operators fulfilling the commutation relations $[\hat{\gamma}_i, \hat{\overline{\gamma}}_j] = \delta_{ij}$ and $[\hat{\gamma}_i, \hat{\gamma}_j] = [\hat{\overline{\gamma}}_i, \hat{\overline{\gamma}}_j] = 0$, and defined the energy functional as
\begin{equation}
    E_s \big( \Phi_s, \overline{\Phi}_s\big) = \int dx \,\overline{\Phi}_s \Big(\hat{H}_{\mathrm{nH}} - \mu + \frac{g}{2} \rho_s  \Big)\Phi_s \, .
\end{equation}. 

They can be written following the Bogoliubov prescription as: 
\begin{equation}
\begin{aligned}
\hat{\gamma_i} =&  \int dx \big[\bar{u}_i(x)  \delta \hat{\Phi}(x) + \bar{v}_i(x) \delta \hat{\overline{\Phi}}(x) \big] \\
\hat{\overline{\gamma_i}} =& \int dx \big[u_i(x) \delta \hat{\overline{\Phi}}(x) + v_i(x)  \delta \hat{\Phi}(x) \big]. 
\end{aligned}
\end{equation}
In order to fulfil the commutation rules, one needs the following orthogonality relation:
\begin{equation}
\int dx \big(\bar{u}_i u_j - \bar{v}_i v_j\big) = \delta_{ij} 
\end{equation}
which is equivalent to the biorthogonality relation in Eq.~(\ref{eq:biorthogonality}), and $\overline{u}$ and $\overline{v}$ can be associated with the left eigenvectors components.

From the commutation relation $\big[\hat{\gamma}_i,~\hat{\mathcal{K}}\big]~=~\epsilon_i \hat{\gamma}_i$ one obtains a BdG operator in the following form (see \cite{supplemental} for the derivation):
\begin{equation}
        \hat{\mathcal{L}}_\mathrm{nH} = 
         \begin{pmatrix}
        \hat{H}_{\mathrm{nH}} -\mu + 2g \rho_s & -g \Phi_s^2 \\
        g(\overline{\Phi}_s)^2 & -\hat{H}^T_{\mathrm{nH}} + \mu - 2g \rho_s 
    \end{pmatrix} \, .
    \label{eq:bdg_quantum}
\end{equation}

\begin{figure*}[!tp]
    \begin{subfigure}[b]{0.3\linewidth}
       \centering
        \begin{overpic}[width=\linewidth]{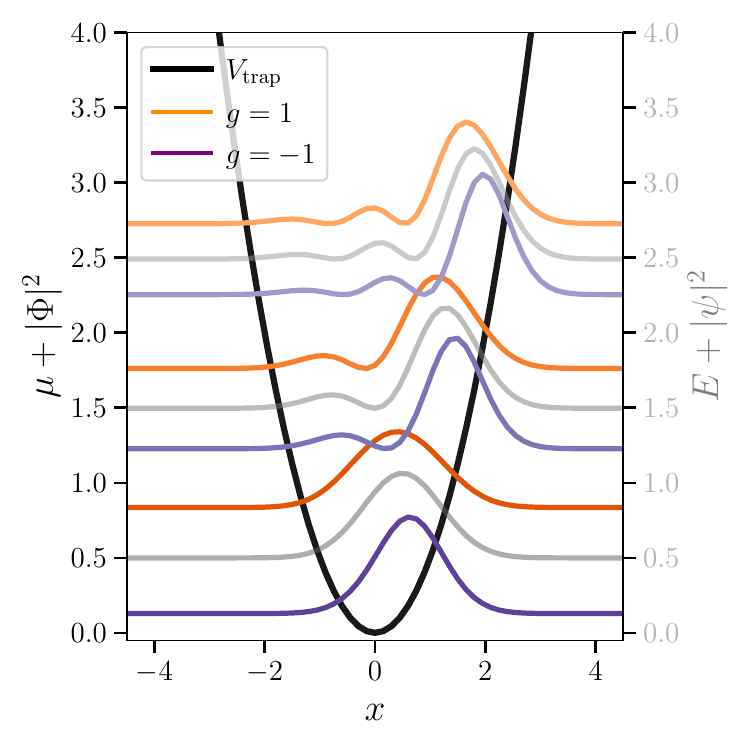}
        \put(-3,142){(a)}
        \end{overpic}
        \phantomcaption
        \label{fig:nonlinear_combined_mt_a}
    \end{subfigure}
     \begin{minipage}[b]{0.3\linewidth}
         \begin{subfigure}[b]{\linewidth}
        \centering
        \begin{overpic}[width=\linewidth]{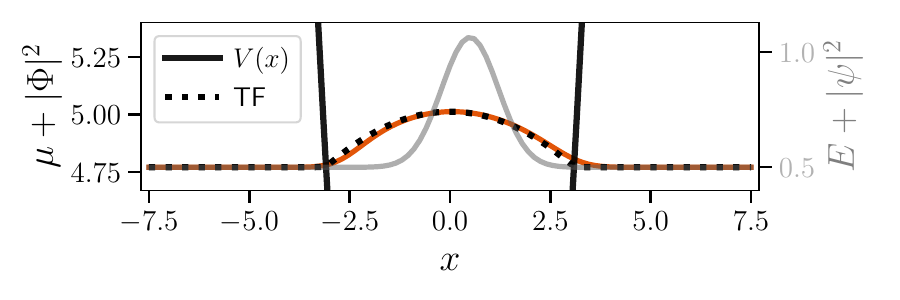}
        \put(-3,55){(b)}
        \end{overpic}
        \phantomcaption
        \label{fig:nonlinear_combined_mt_b}
    \end{subfigure}\\[-0.4cm]
    \begin{subfigure}[b]{\linewidth}
        \centering
        \begin{overpic}[width=\linewidth]{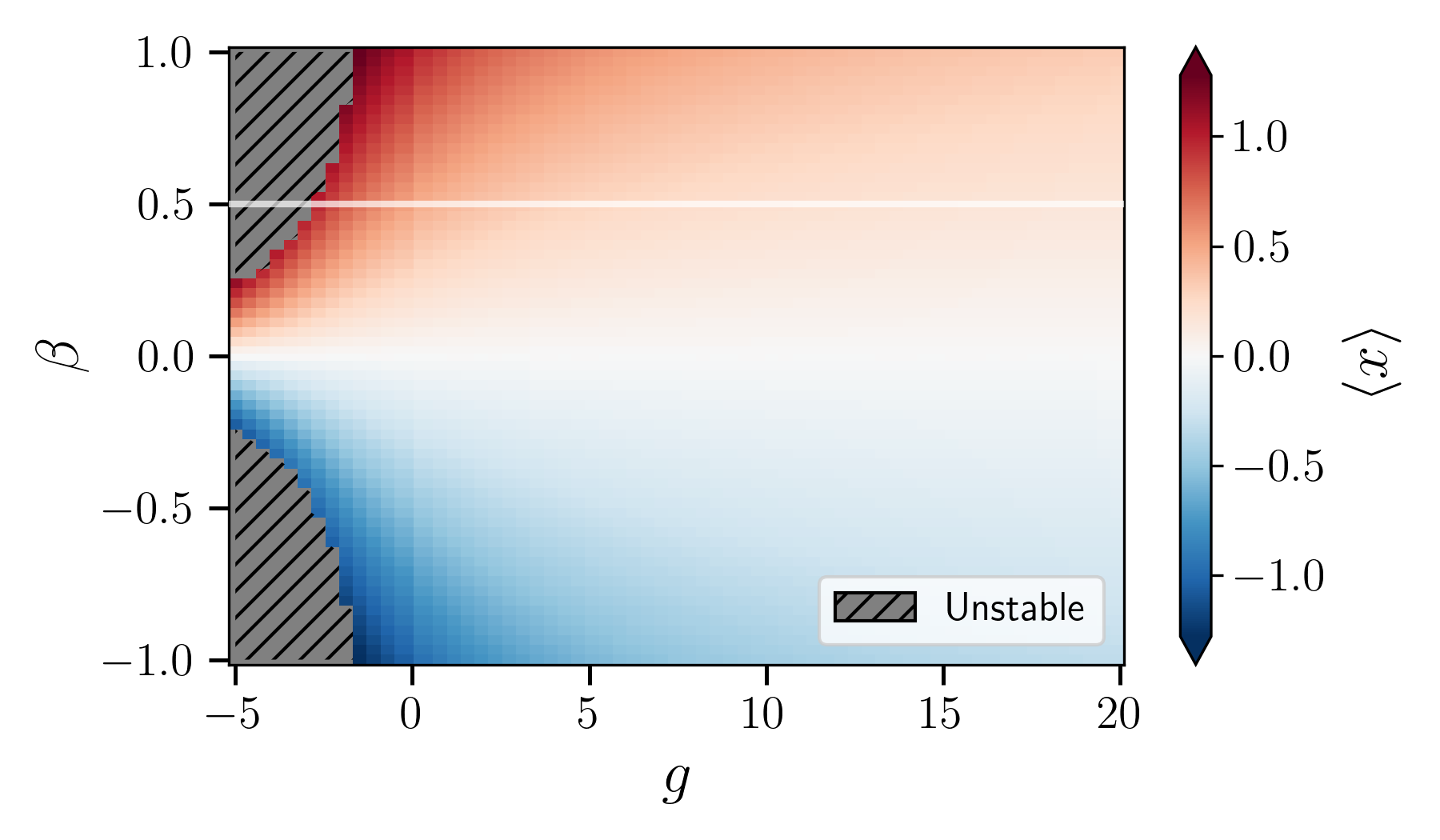}
        \put(-3,80){(c)}
        \end{overpic}
        \phantomcaption
        \label{fig:nonlinear_combined_mt_c}
    \end{subfigure}
    \end{minipage}
    \begin{minipage}[b]{0.28\linewidth}
    \begin{subfigure}[b]{\linewidth}
        \centering
        \begin{overpic}[width=\linewidth]{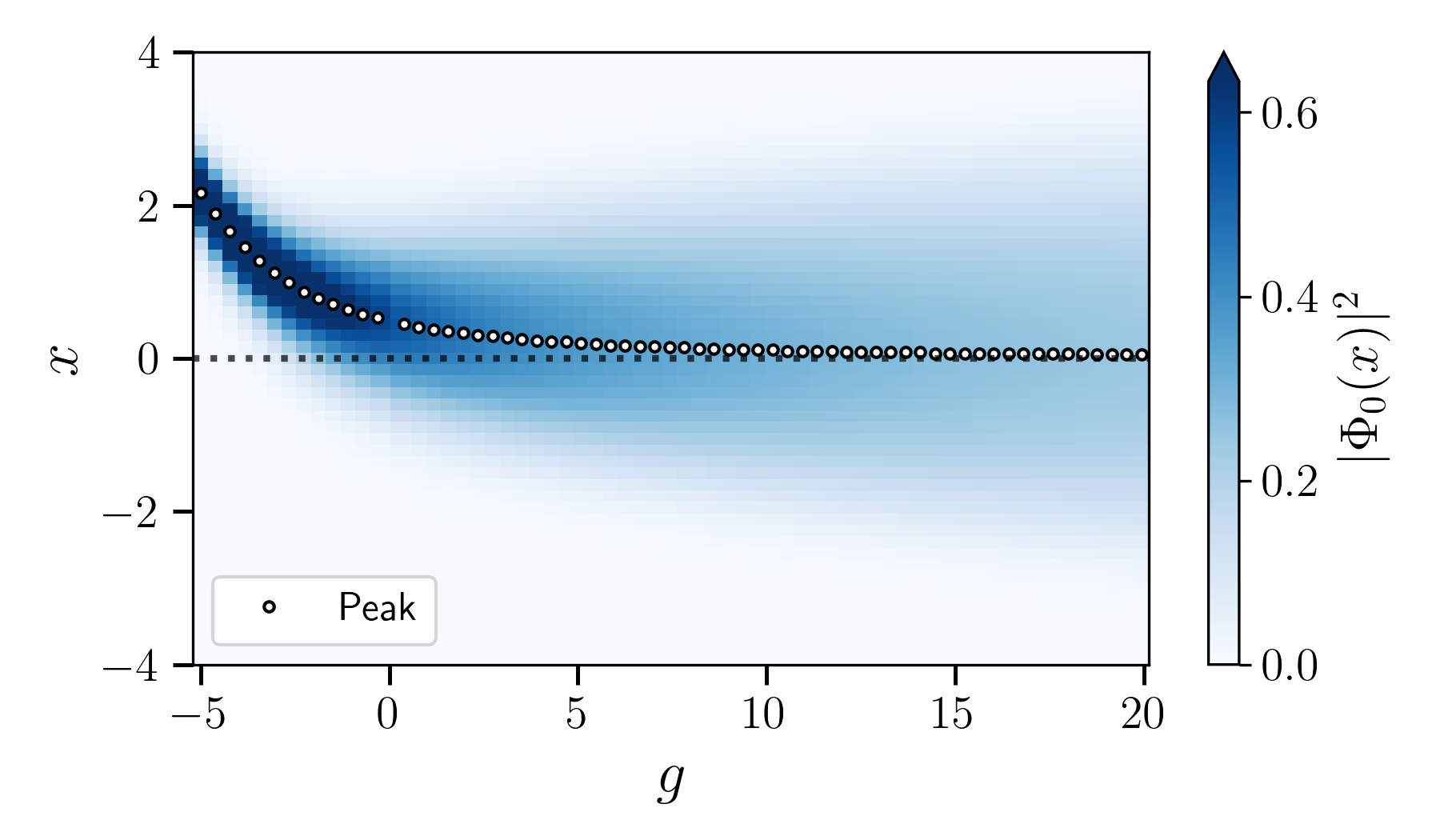}
        \put(-3,72){(d)}
        \end{overpic}
        \phantomcaption
        \label{fig:nonlinear_combined_mt_d}
    \end{subfigure}\\[-0.9cm]
    \begin{subfigure}[b]{\linewidth}
        \centering
        \begin{overpic}[width=\linewidth]{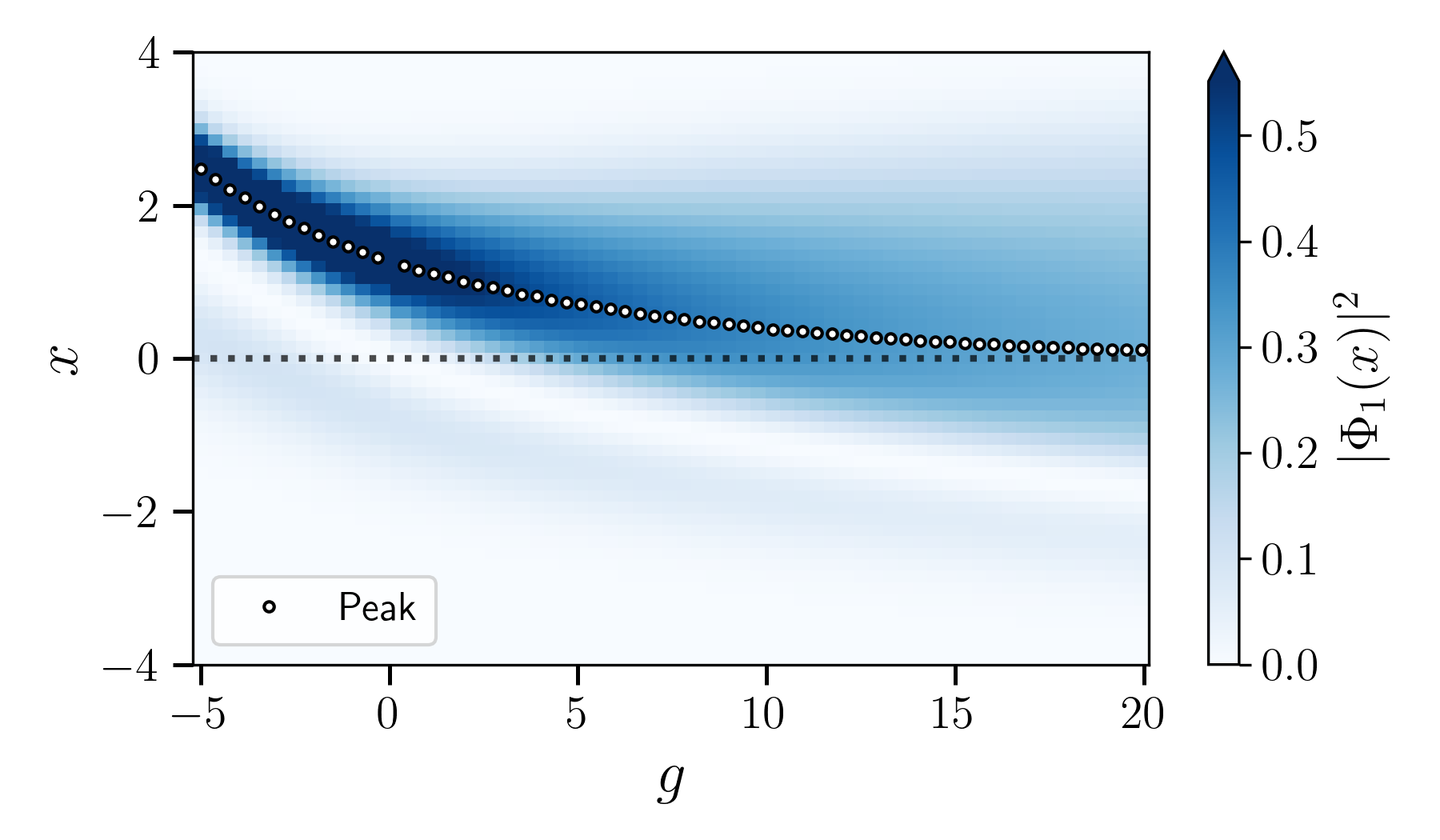}
        \put(-3, 72){(e)}
        \end{overpic}
        \phantomcaption
        \label{fig:nonlinear_combined_mt_e}
    \end{subfigure}
    \end{minipage}\\[-0.4cm]
    \caption{ \small \justifying Nonlinear modes. (a) First three solutions of the nonlinear eigenvalue problem for positive and negative interaction strengths $g=\pm 1$, in orange and purple curves respectively, and the imaginary gauge field strength $\beta = 0.5$. The grey curve corresponds to the linear case ($g=0$). The thick black curve shows the confining harmonic potential. 
    (b) Position expectation value $\langle x \rangle$ of the nonlinear ground state as a function of $\beta$ and $g$. The grey hatched region indicates the dynamically unstable regime, where the BdG spectrum acquires complex eigenvalues. (c) Comparison between the ground state mode (orange) and the Thomas-Fermi solution (dashed black) for $g=20$ and $\beta=0.5$. The grey curve corresponds to the linear case ($g=0$).
    (d,e) Probability density of the (d) ground and (e) first excited state for varying $g$ and a fixed $\beta=0.5$ indicated by the white line in panel (b). The dotted black line marks $x=0$, and the white dots represent the interpolated maximum value.
    }
    \label{fig:nonlinear_combined_mt} 
\end{figure*} 

Note that the imaginary gauge field has an opposite sign in the diagonal blocks of the BdG operator [c.f. Eq.~(\ref{eq:bdg_operator_mt})]. For $\big[\hat{\overline{\gamma}}_i,~\hat{\mathcal{K}}\big]~=~-\epsilon_i \hat{\overline{\gamma}}_i$, one obtains similar equations for negative energies with $(u, v)^T\rightarrow(\overline{v}, \overline{u})^T$. In contrast to Eq. (\ref{eq:bdg_operator_mt}), we can perform a similarity transformation reducing to the Hermitian ($\beta \rightarrow 0$) BdG problem $\mathcal{S'} \mathcal{L}_{\mathrm{nH}} \mathcal{S'}^{-1} = \mathcal{L}_{\mathrm{H}}$, with $\mathcal{S'} = \mathrm{diag}(S, S^{-1})$, hence the eigenvalues remain unchanged. 

In this framework, many quantum observables built from the Bogoliubov solutions are insensitive to the non-Hermiticity of the Hamiltonian. As an example, consider the noncondensate density obtained from the ensemble average:
\begin{equation}
\begin{aligned}
    \langle\delta \hat{\overline{\Phi}}\, \delta \hat{\Phi}\rangle
    &= \sum_{ij} \big\langle \big(\bar{u}_i \hat{\bar{\gamma}}_i
       - v_i \hat{\gamma}_i\big)
       \big(u_j \hat{\gamma}_j - \bar{v}_j \hat{\bar{\gamma}}_j\big)
       \big\rangle \\
    &= \sum_i \big(\bar{u}_i u_i + \bar{v}_i v_i\big)
       \langle \hat{\bar{\gamma}}_i \hat{\gamma}_i \rangle
       + \bar{v}_i v_i \, ,
\end{aligned}
\end{equation}
where we used the mode expansions
\begin{equation}
\begin{aligned}
    \delta \hat{\Phi}(x) &= \sum_i u_i(x)\, \hat{\gamma}_i
        - \bar{v}_i(x)\, \hat{\bar{\gamma}}_i \, , \\
    \delta \hat{\overline{\Phi}}(x) &= \sum_i \bar{u}_i(x)\, \hat{\bar{\gamma}}_i
        - v_i(x)\, \hat{\gamma}_i \, ,
\end{aligned}
\end{equation}
together with the biorthogonal commutation relations $[\hat{\gamma}_i, \hat{\bar{\gamma}}_j] = \delta_{ij}$, and $\langle \hat{\bar{\gamma}}_i \hat{\gamma}_i \rangle$ reduces to a Bose-Einstein distribution. Since only the biorthogonal bilinears $\bar{u}_i u_i$ and $\bar{v}_i v_i$ appear in the above expressions, the exponential envelopes generated by the imaginary gauge field cancel identically; static observables of this type do not experience the skin effect.

The situation is different for transport quantities. Noether's theorem yields the continuity equation \cite{supplemental}:
\begin{equation}
    \partial_t \big(\overline{\Phi}\Phi\big) + \partial_x j = 0 \, ,
\end{equation}
with a current modified with respect to the Hermitian case,
\begin{equation}
    j = \frac{1}{2i} \Big[ (\partial_x \overline{\Phi})\Phi
        - \overline{\Phi}\, (\partial_x\Phi)
        - 2\beta\, \overline{\Phi}\Phi \Big] \, .
\end{equation}
This Noether current differs from the hydrodynamic one obtained from the Madelung transformation, $j_{\mathrm{hydro}} \equiv n\mathrm{v} = n\,\partial_x\vartheta$, which instead satisfies \cite{supplemental}
\begin{equation}
    \partial_t n + \big(\partial_x - 2\beta\big)\, j_{\mathrm{hydro}} = 0 \, .
\end{equation}
Here the term $2\beta\, j_{\mathrm{hydro}}$ acts as a local source: the density $n$ is not conserved in the usual sense, and $j_{\mathrm{hydro}}$ is not a Noether current. It is the biorthogonal density $\overline{\Phi}\Phi$, rather than $n$, that obeys a conservation law.

{\it NHSE manifestation ---} As the first step, we find stationary solutions of Eq. (\ref{eq:gross_pita}) corresponding to real values of $\mu$ numerically. This is equivalent to finding a root of $F(\Phi(x), \mu) = 0$ \cite{Panu2026}, taking the linear solutions in the limit $g \rightarrow 0$ as an initial guess. To this end, we discretise the Hamiltonian on a one-dimensional lattice and incorporate the imaginary gauge field via an imaginary Peierls substitution, in analogy to the standard Peierls treatment for real magnetic fields \cite{Peierls1933, HN1996}. The resulting one dimensional tight-binding Hamiltonian is:
\begin{equation}
    \hat{H}_{\mathrm{tb}} =  \sum_i (V_i -2t) \, \hat{b}^{\dagger}_i \hat{b}_i + t_+ \, \hat{b}^{\dagger}_{i} \hat{b}_{i+1}  + t_- \, \hat{b}^{\dagger}_{i+1} \hat{b}_i  ,
    \label{eq:tb_ham}
\end{equation}
where $\hat{b}_i^{\dagger}$ and $\hat{b}_i$ are bosonic creation and annihilation operators on the site $i$. The hopping parameter is $t=1/2a^2$, where $a$ is the discretisation parameter (lattice constant), and 
\begin{equation}
    t_{\pm} = -t e^{\pm a \beta }
\end{equation}
define the non-reciprocal hopping terms originating from the imaginary gauge field. 

The expectation value on the vacuum for $\hat{H}_{tb}$ is substituted in Eq. (\ref{eq:bdg_operator_mt}) instead of $H_{\mathrm{nH}}$ and similarly the remaining quantities in Eq. (\ref{eq:gross_pita}) and $\mathcal{L}$ result in on-site terms, \textit{i.e.} diagonal matrices in the lattice space. 
In Fig. \ref{fig:nonlinear_combined_mt_a} we present the stationary solutions of Eq. \ref{eq:gross_pita} for the ground, the first, and second excited eigenstates in a harmonic trap $V_{\mathrm{trap}}(x) = x^2 /2$ (in dimensionless units \cite{supplemental}), which is the standard choice for studying collective excitations of trapped condensates \cite{Stringari1996, Dalfovo1999}. We observe the localisation of all the eigenfunctions on the right boundary of the confining potential, as expected, arising from the imaginary gauge term in different regimes: linear skin effect ($g=0$) and the skin effect in the presence of repulsive ($g>0$) or attractive ($g<0$) nonlinearity. One can observe the broadening (narrowing) of the nonlinear state peak \cite{Dalfovo1999} with respect to the linear case for the repulsive (attractive) interaction. 

In the Thomas-Fermi limit, $\mu \gg \omega$ (obtained when $g \gg 1$), the solutions for Eq. (\ref{eq:gross_pita}) can be approximated as: 
\begin{equation}
     |\Phi(x)|^2 \simeq \frac{2\mu + \beta^2 -x^2}{2g} \theta(2\mu + \beta^2 -x^2)
\end{equation}
where $\mu = \big[(3g/2)^{2/3}-\beta^2\big]/2$ is found from the normalisation condition $\int|\Phi(x)|^2dx=1$. The regime under which we discard the gauge term is $\beta \xi_h \ll 1$, where $\xi_h$ is the healing length \cite{pitaevskii2016} defined as $\xi_h = (3g)^{-1/3}$ \footnote{From an experimental configuration one can retrieve the dimensionless interaction strength from $g= \frac{g_e N}{\hbar \omega l}$, where $N$ is the number of atoms and $l = \sqrt{\hbar/m\omega}$. See the supplemental material \cite{supplemental} for more details.}.
Figure \ref{fig:nonlinear_combined_mt_b} shows that the numerical solution matches the Thomas-Fermi solution very well. In addition, we can see that the localisation strength due to the linear skin effect is weakened, especially when compared to the linear case ($g=0$) shown as the grey curve.

\begin{figure*}[!tp]
\begin{minipage}[b]{0.25\linewidth}
    \begin{subfigure}[b]{\linewidth}
        \centering
        \begin{overpic}[width=\linewidth]{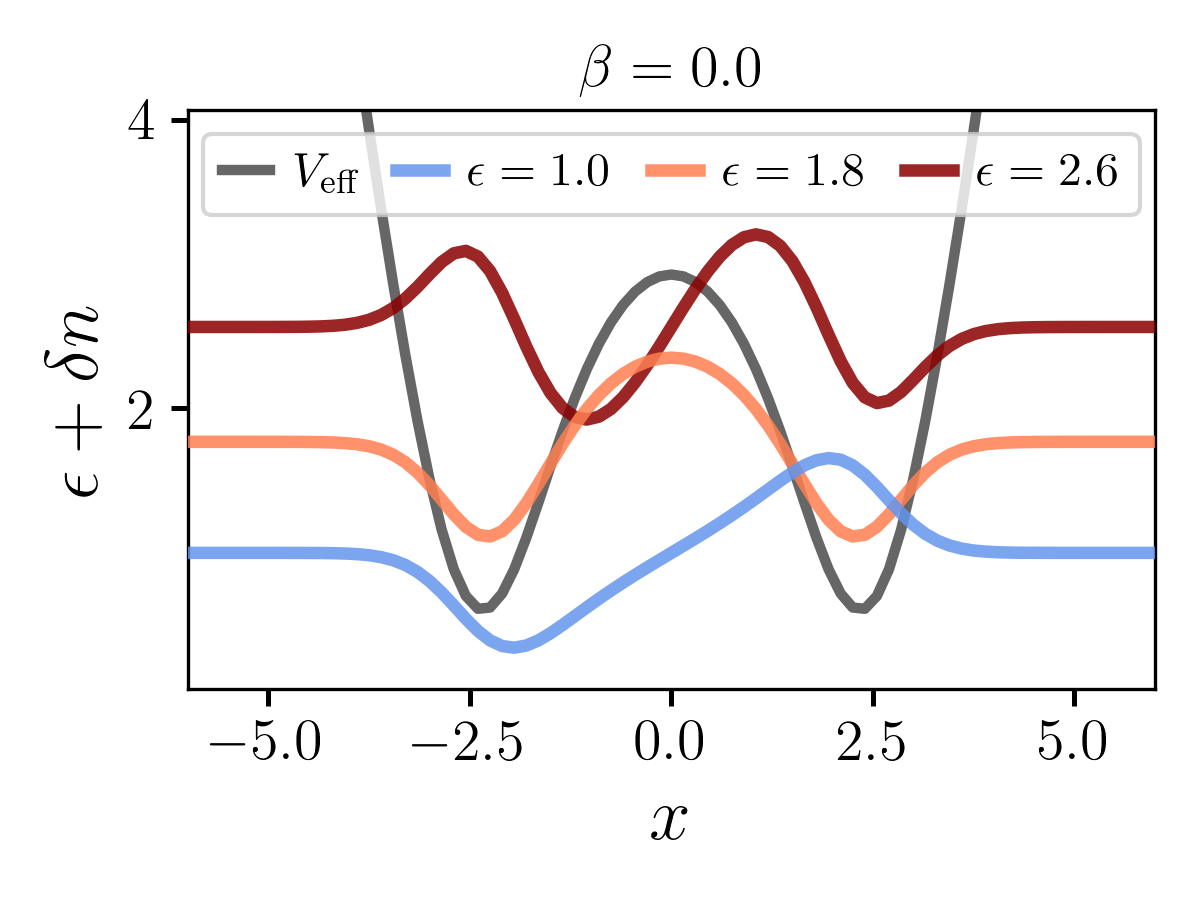}
        \put(-1,85){(a)}
        \end{overpic}
        \phantomcaption
        \label{fig:bdg_combined_a}
    \end{subfigure}\\[-0.5 cm]
    \begin{subfigure}[b]{\linewidth}
        \centering
        \begin{overpic}[width=\linewidth]{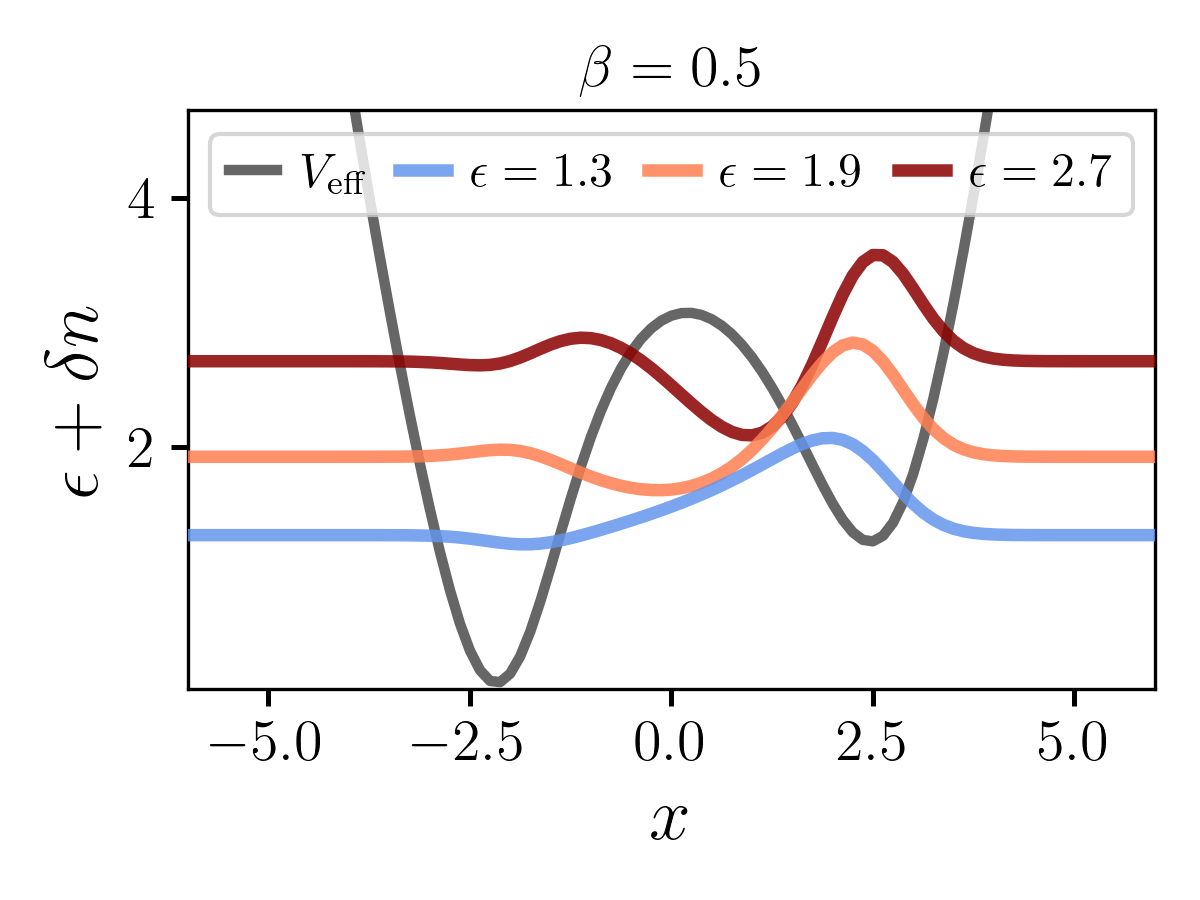}
        \put(-1,85){(b)}
        \end{overpic}
        \phantomcaption
        \label{fig:bdg_combined_b}
    \end{subfigure}
\end{minipage}
\begin{minipage}[b]{0.23\linewidth}
    \begin{subfigure}[b]{\linewidth}
        \centering
        \begin{overpic}[width=\linewidth]{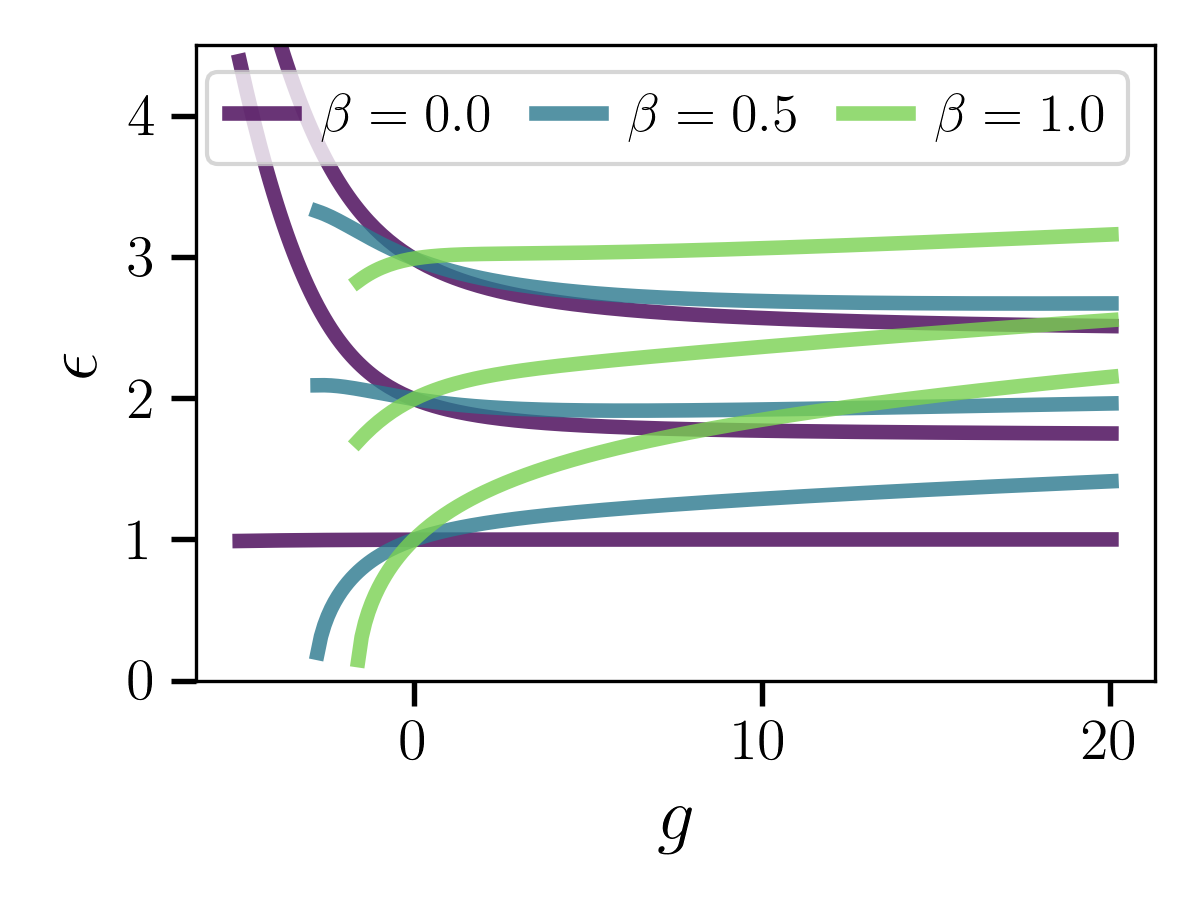}
        \put(-1,85){(c)}
        \end{overpic}
        \phantomcaption
        \label{fig:bdg_combined_c}
    \end{subfigure}\\[-0.24 cm]
    \begin{subfigure}[b]{\linewidth}
        \centering
        \begin{overpic}[width=\linewidth]{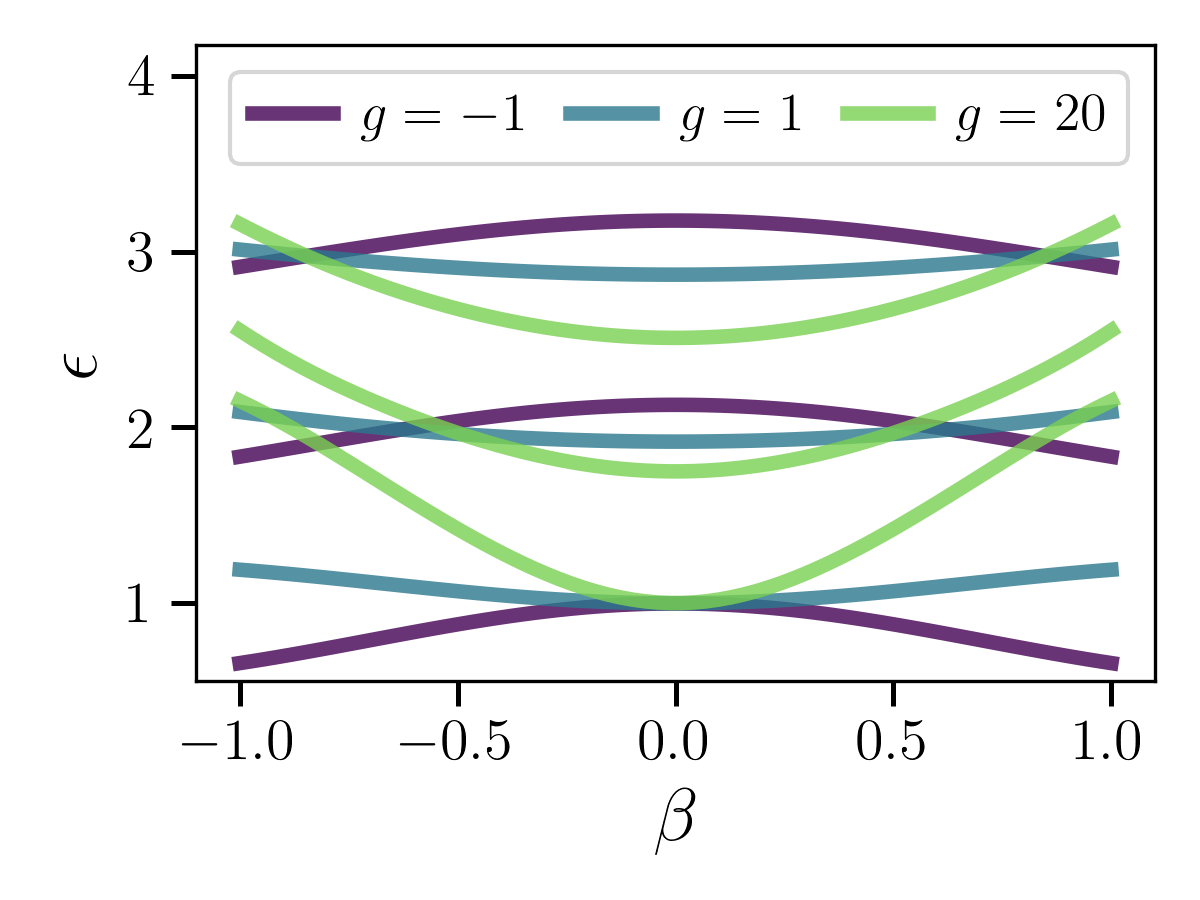}
        \put(-1,85){(d)}
        \end{overpic}
        \phantomcaption
        \label{fig:bdg_combined_d}
    \end{subfigure}
\end{minipage}
\begin{minipage}[b]{0.25\linewidth}
    \begin{subfigure}[b]{\linewidth}
        \centering
        \begin{overpic}[width=\linewidth]{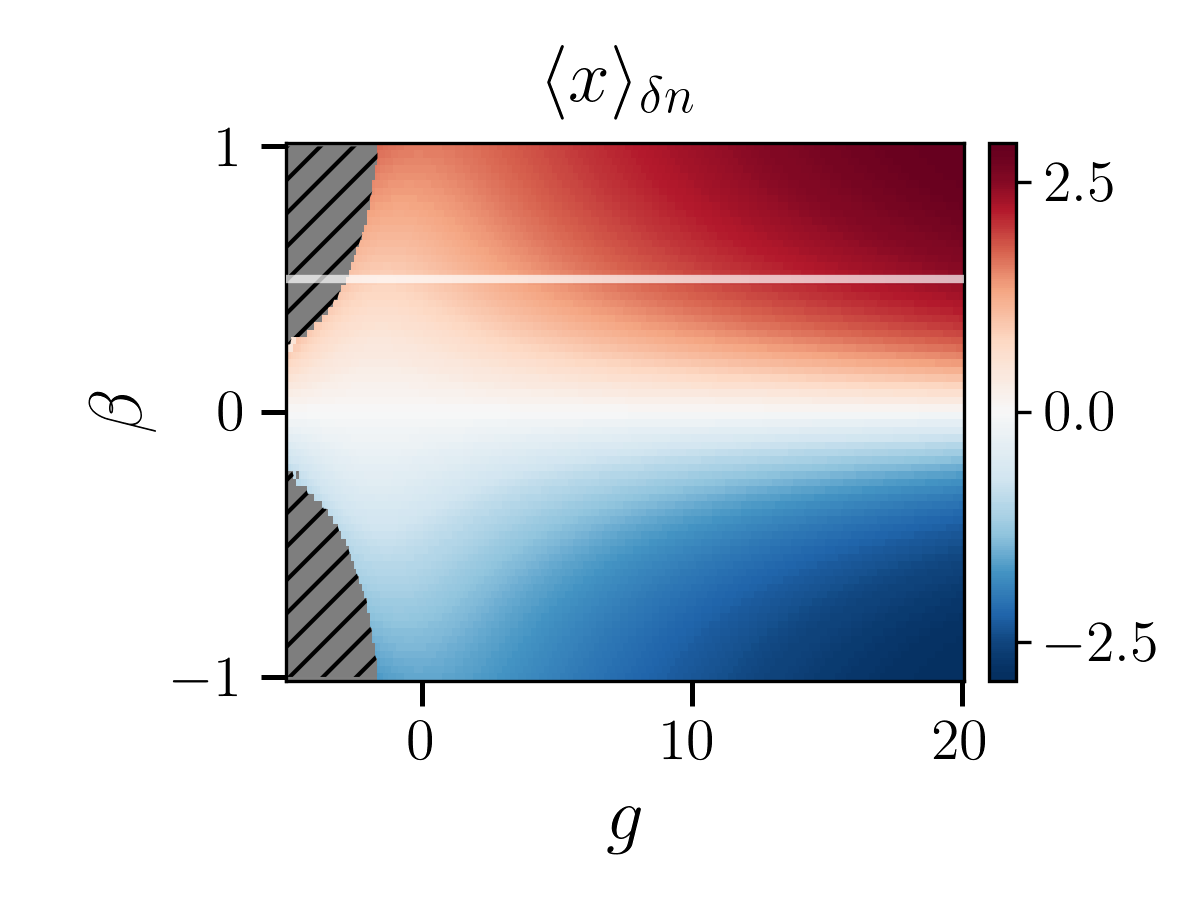}
        \put(+1,85){(e)}
        \end{overpic}
        \phantomcaption
        \label{fig:bdg_combined_e}
    \end{subfigure}\\[-0.5 cm]
    \begin{subfigure}[b]{\linewidth}
        \centering
        \begin{overpic}[width=\linewidth]{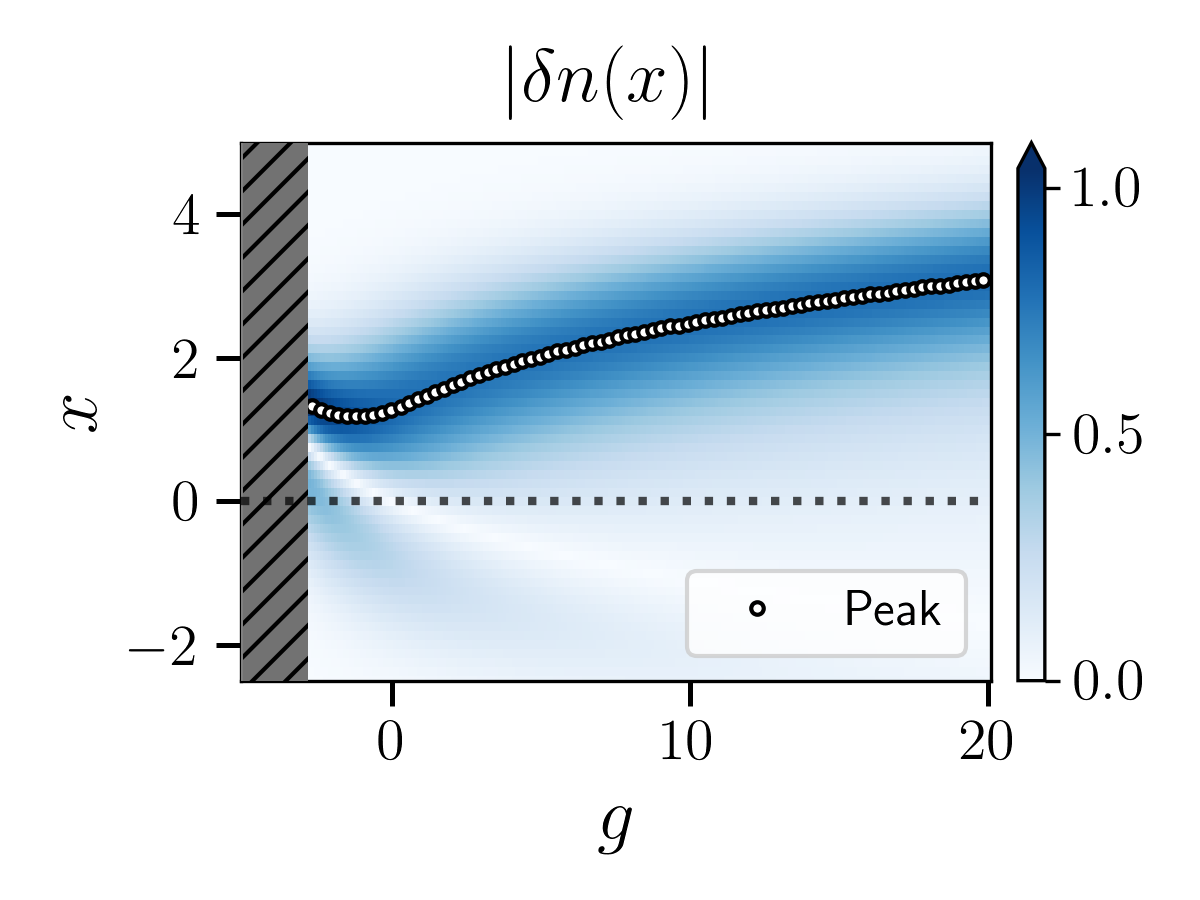}
        \put(+1,85){(f)}
        \end{overpic}
        \phantomcaption
        \label{fig:bdg_combined_f}
    \end{subfigure}
\end{minipage}
\begin{minipage}[b]{0.25\linewidth}
    \begin{subfigure}[b]{\linewidth}
        \centering
        \begin{overpic}[width=\linewidth]{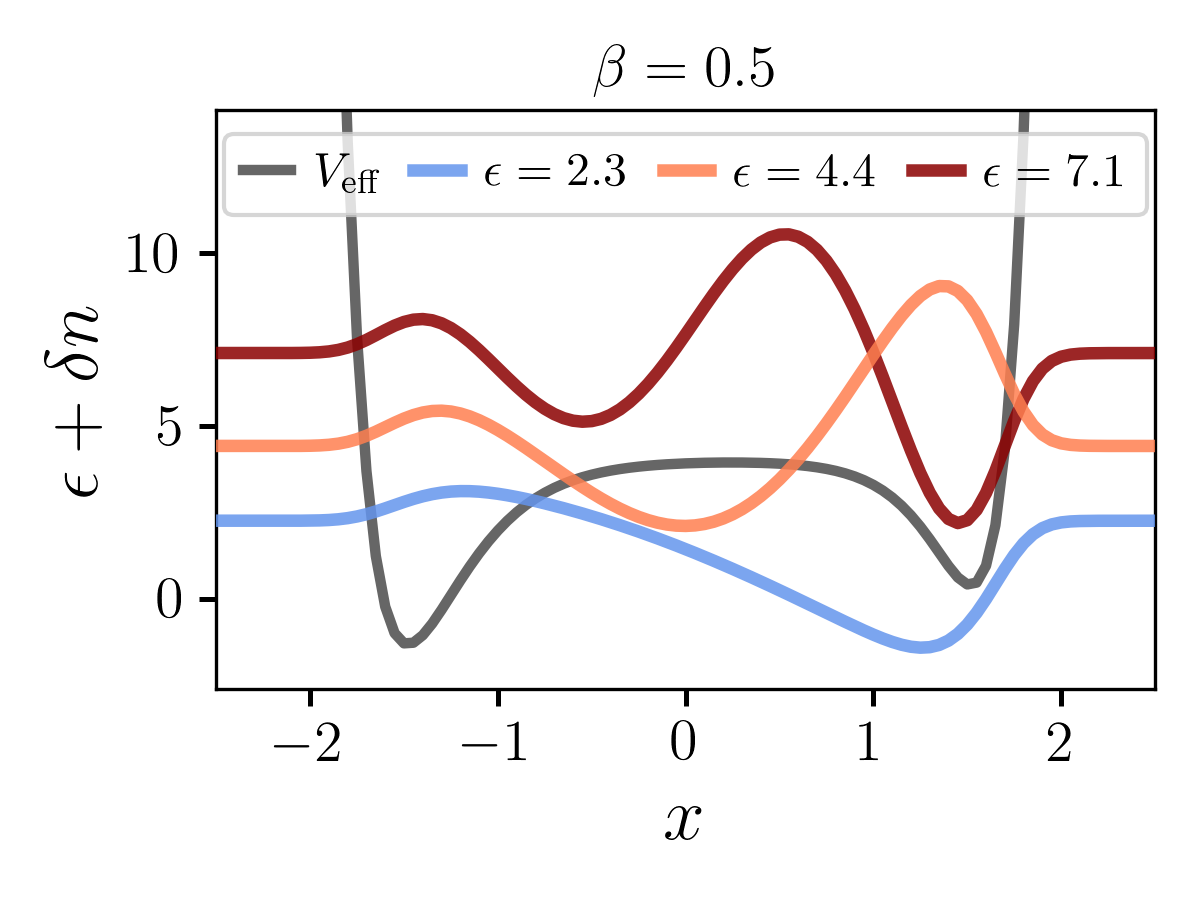}
        \put(+1,85){(g)}
        \end{overpic}
        \phantomcaption
        \label{fig:bdg_combined_g}
    \end{subfigure}\\[-0.5 cm]
    \begin{subfigure}[b]{\linewidth}
        \centering
        \begin{overpic}[width=\linewidth]{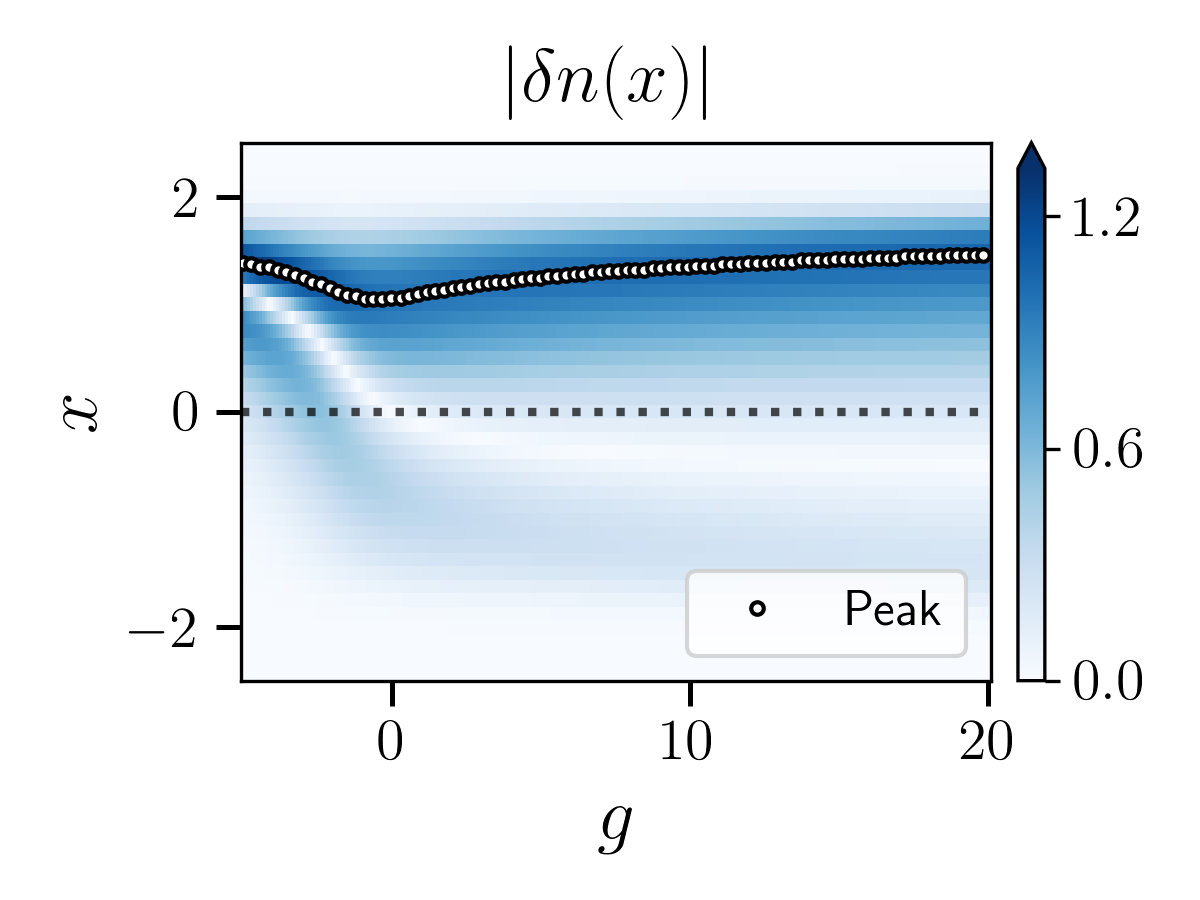}
        \put(+1,85){(h)}
        \end{overpic}
        \phantomcaption
        \label{fig:bdg_combined_h}
    \end{subfigure}
\end{minipage}\\[-0.4cm]
    \caption{ \small \justifying Bogoliubov-de Gennes modes. (a,b) The three lowest-order modes of density perturbation, together with the effective potentials $V_{\mathrm{eff}}(x)$, obtained by diagonalising the BdG operator $\mathcal{L}$. Parameters are: $g=10$, (a) $\beta=0$, and (b) $\beta=0.5$. (c,d) Eigenvalues of the first three excited BdG modes, as a function of (c) $g$ and (d) $\beta$. (e) Position expectation value for the first excited BdG state as a function of $g$ and $\beta$. The grey hatched regions mark dynamically unstable solutions. (h) Density of the BdG mode at $\beta=0.5$ marked by the white line in panel (e). Black dashed line marks $x=0$, and white dots highlight the peak of $|\delta n|$. For panels (a-f), the confinement is due to a harmonic potential $V_{\mathrm{trap}}(x) = x^2/2$. (g,h) Same as panels (a) and (f) but for a super-Gaussian trapping potential $V_{\mathrm{trap}}(x) = V_0 \mathrm{exp}\{-(x/\sigma)^m\}$ with $V_0 = 100$, $\sigma=2$, and $m=16$.
    }
    \label{fig:bdg_combined}
\end{figure*}

We can characterise the effect of the interaction strength by plotting the position expectation value 
\begin{equation}
    \langle x \rangle = \int x \, |\Phi_s(x)|^2 \, dx \, ,
    \label{eq:x_mean}
\end{equation}
with $\int |\Phi_s(x)|^2 \, dx = 1$, for varying $g$ and $\beta$ (see Fig. \ref{fig:nonlinear_combined_mt_c}).
Figures \ref{fig:nonlinear_combined_mt_d} and \ref{fig:nonlinear_combined_mt_e} demonstrate the behaviour of the density of the ground and first excited states at a finite $\beta$ and varying $g$, showing that the NHSE disappears for large $g$. This behaviour is connected to the fact that, in the interaction-dominated Thomas-Fermi limit, the kinetic term in the Hamiltonian is no longer relevant and the imaginary gauge field does not play a role.

As discussed above, repulsive interactions ``screen'' the NHSE, while attractive interactions have the opposite effect: within the dynamically stable regime they enhance the position displacement due to NHSE. The instability of the solutions is signalled by a non-vanishing imaginary part of the $\mathcal{L}$ eigenvalues $\mu$. 

We note that these key results are general and are not limited to the case of harmonic confinement, as demonstrated in SM (\cite{supplemental}) for a super-Gaussian potential defined as $V_{\mathrm{trap}}(x) = V_0 (1 - \mathrm{exp}\{-(x/\sigma)^m\})$ (see Fig. \ref{fig:nonlinear_combined}). 

Similarly to the nonlinear solutions, we can now investigate the BdG excitations and how their spatial density distribution behaves when the spatial profile of the ground state is affected by the NHSE. In Fig. \ref{fig:bdg_combined} we present the results obtained by diagonalising the BdG operator in Eq. (\ref{eq:bdg_operator_mt}), where we again replaced $H_{\mathrm{nH}}$ with the discretised Hamiltonian Eq.~(\ref{eq:tb_ham}). In Figs. \ref{fig:bdg_combined_a} and \ref{fig:bdg_combined_b}, we compare the density [see Eq. (\ref{eq:density_phase})] of the first three BdG modes ordered in terms of their respective eigenvalues in the cases of a vanishing and finite imaginary gauge term $\beta$, respectively, in the presence of a harmonic trapping potential.  We also plot the effective potential defined as $V_{\mathrm{eff}}(x)~=~V_{\mathrm{trap}}(x)~+~2g|\Phi_0(x)|^2-\mu$ as a reference. We do not show the solutions for negative energies since they are easily obtained from the particle-hole symmetry discussed above.

\begin{figure*}[!tp]
    \centering
    \begin{subfigure}[b]{0.45\linewidth}
        \centering
        \begin{overpic}[width=\linewidth]{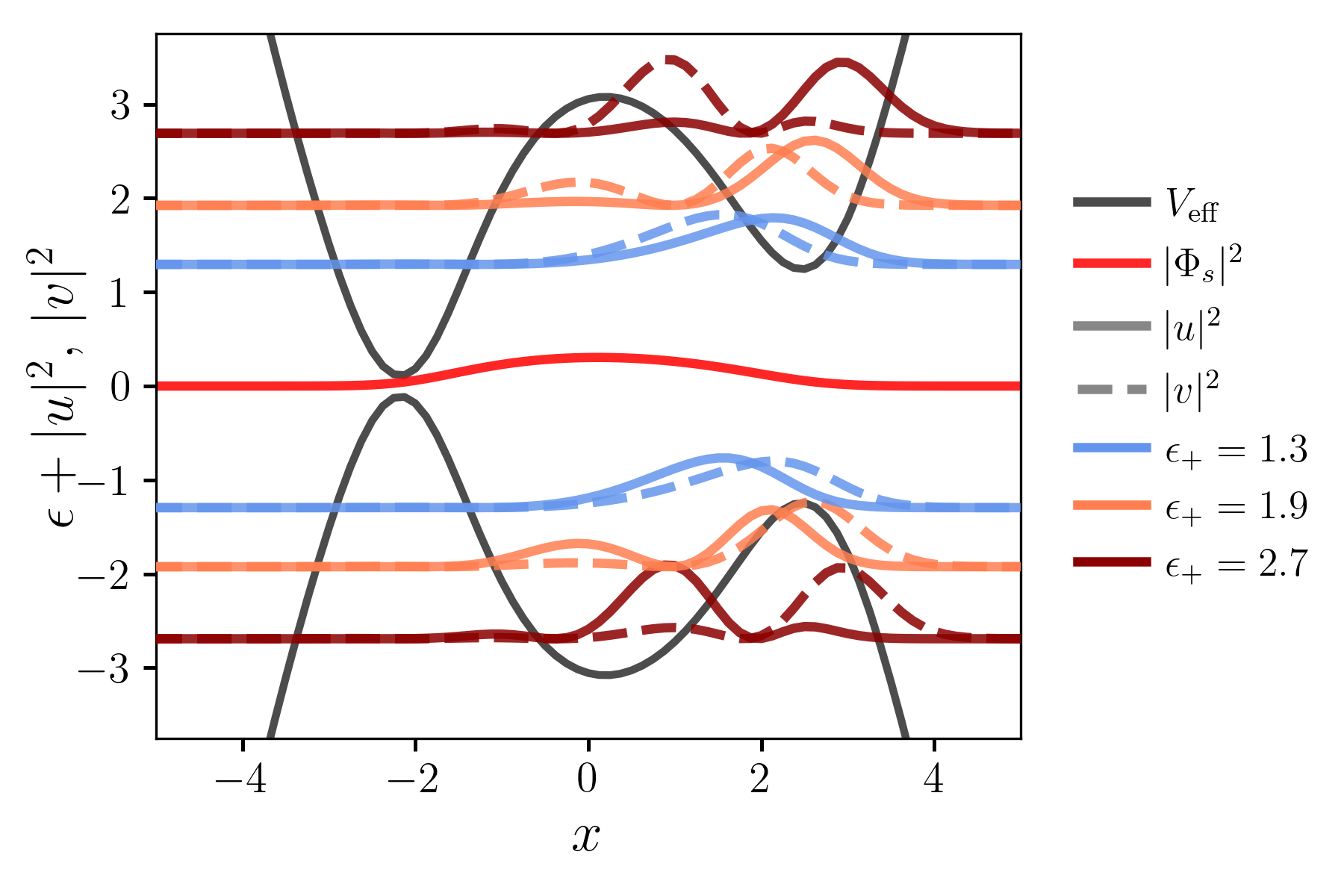}
        \put(+1,140){(a)}
        \end{overpic}
        \phantomcaption
        \label{fig:flip_comp_a}
    \end{subfigure}
    \begin{subfigure}[b]{0.45\linewidth}
        \centering
        \begin{overpic}[width=\linewidth]{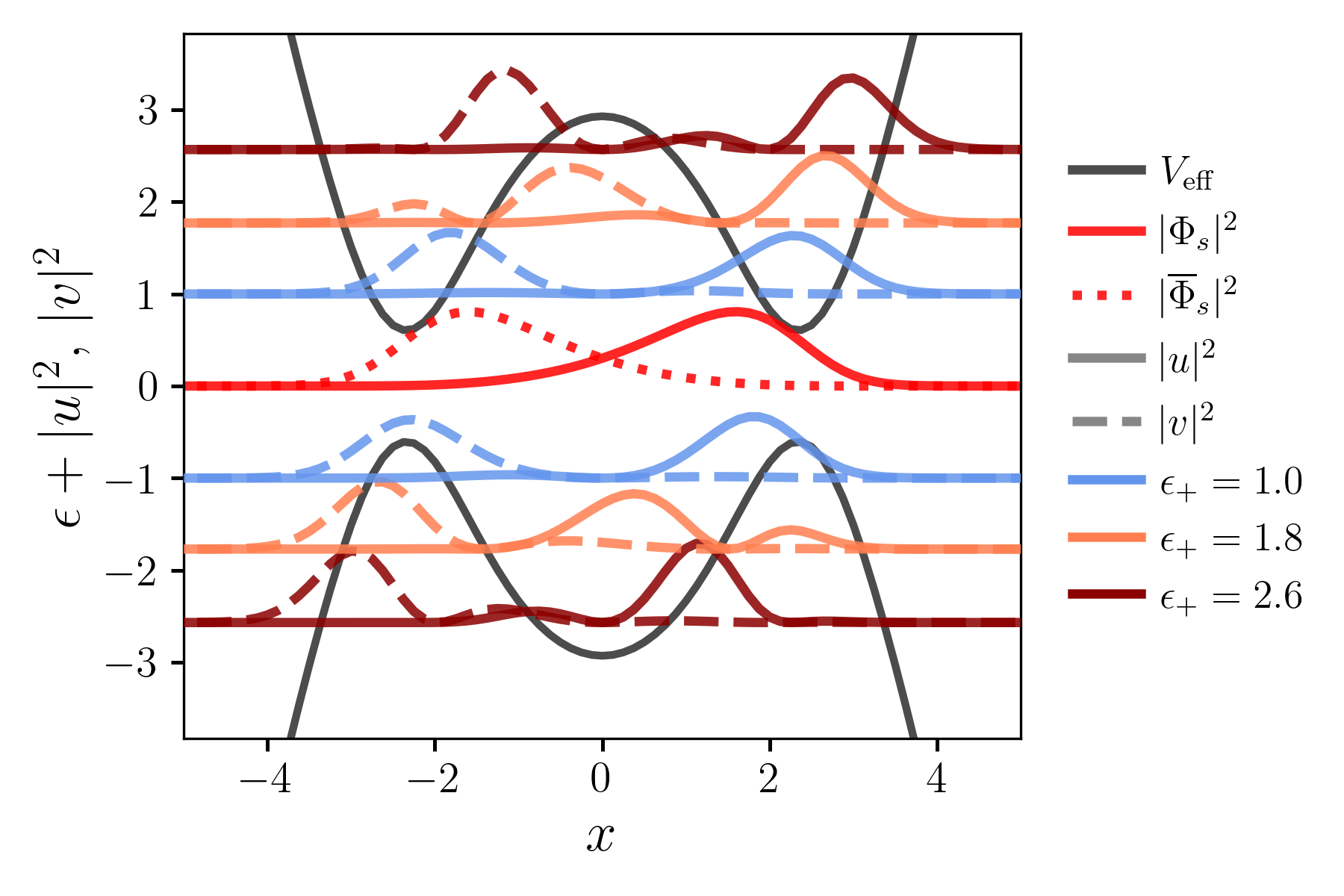}
        \put(+1,140){(b)}
        \end{overpic}
        \phantomcaption
        \label{fig:flip_comp_b}
    \end{subfigure}\\[-0.4cm]
    \caption{ \small \justifying (a,b) Eigenfunction components $u$ (solid line) and $v$ (dashed line) of the BdG operator $\mathcal{L}$ plotted together with the effective potentials $\pm V_{\mathrm{eff}}(x)$ (solid black line). The eigenfunctions components $u$ and $v$ are independently normalised for visualisation purposes. The parameters are $g=10$ and $\beta=0.5$. Only the components with positive eigenvalues $\epsilon_+$ are shown as the negative ones are their symmetric counterparts $\epsilon_- = - \epsilon_+$. (a) Solutions of Eq. (\ref{eq:bdg_operator_mt}). The ground state of the nonlinear problem in Eq. \ref{eq:gross_pita} is shown as a red line. (b) Solutions of Eq.(\ref{eq:bdg_quantum}). Ground states of the nonlinear set of Eqs. (\ref{eq:gross_pita_quantum}) are represented by a solid and dashed line.}
    \label{fig:flip_comp}
\end{figure*}

The three lowest-order BdG modes are known as the dipole, monopole (or breathing), and quadrupole excitations \cite{Stringari1996, pitaevskii2016}. In particular, we focus on the dipole mode because of its special role among the collective excitations of a harmonically trapped condensate. In the Hermitian limit, its eigenvalue is pinned to the trap frequency irrespective of the interaction strength \cite{Dobson1994} -- an exact result protected by Kohn's theorem \cite{Kohn1961}, which holds as long as the external potential is purely harmonic. The introduction of an imaginary gauge field breaks the conditions underlying Kohn's theorem, lifting the protection of the dipole mode. Its eigenvalue shifts away from the trap frequency and acquires a nontrivial dependence on the interaction strength $g$ (see Fig. \ref{fig:bdg_combined_c}). Moreover, Fig. \ref{fig:bdg_combined_d} reveals that this shift grows with the strength of the NHSE, making the dipole mode frequency a direct quantitative probe of the latter. One could expect that this drift would vanish at high values of $g$, \textit{i.e.} in the TF limit, since the modified kinetic term should play a minor role. The obtained result suggests that this is not the case: as we already observed in Eq. (\ref{eq:hydrodynamic_TF}), even in this limit $\beta$ will affect both the spatial distribution of the modes and the eigenvalues properties. 

In Fig. \ref{fig:bdg_combined_e}, we show the position expectation values of the first excited state. In addition, in Fig. \ref{fig:bdg_combined_f} we scan the density of the corresponding solutions for a fixed $\beta$ and varying $g$. We observe that, in stark contrast with the solutions of Eq. (\ref{eq:gross_pita}), increasing the nonlinearity enhances the asymmetry of the spatial distribution of the linear excitation modes. This can be understood in the Thomas-Fermi limit: since the solutions are confined in the range $|x|<\sqrt{2\mu_{\tiny{\mathrm{TF}}}}\sim g^{1/3}$ \cite{Gao2020}, the imaginary gauge field will still shift the distribution to a new effective confining potential whose boundary position is proportional to $g$. For this reason, at large values of $g$, the NHSE is inherited by the BdG excitations, even if it does not affect the eigenfunctions of the confining potential. To make this point more evident, Figs. \ref{fig:bdg_combined_g} and \ref{fig:bdg_combined_h} show the solutions for a super-Gaussian trapping potential defined above. In this case, there is no enhancement of the localisation given the hard wall-like confinement.

Finally, we compare the manifestations of NHSE predicted within the classical and the second-quantised framework. Figure~\ref{fig:flip_comp} shows the eigenvectors and the spectra
obtained from the two definitions of $\hat{\mathcal{L}}$: in panel~\ref{fig:flip_comp_a} we follow the prescription of Eq.~(\ref{eq:bdg_operator_mt}), as in Fig.~\ref{fig:bdg_combined}, whereas in panel~\ref{fig:flip_comp_b} we solve the problem for $\mathcal{L}$ as defined in Eq.~(\ref{eq:bdg_quantum}). The two approaches yield different sets of eigenvectors and eigenvalues. In particular, the latter approach exhibits no shift of the dipole-mode frequency away from Kohn's protected modes~\cite{Kohn1961}, which sit at $\epsilon = \pm 1$.
This is not surprising as $\rho_s$, which enters Eq.~(\ref{eq:bdg_quantum}) is not influenced by the NHSE.
Moreover, whereas the first case requires computing the left eigenvectors in order to enforce the biorthogonality condition~(\ref{eq:biorthogonality}), the solutions of Eq.~(\ref{eq:bdg_quantum}) automatically provide all the eigenvectors needed for it. We note how the $u$ and $v$ components ``feel'' the skin effect differently in the two cases. Indeed, in Fig.~\ref{fig:flip_comp_b} the negative-energy components $v = \overline{u}$ and $u = \overline{v}$ are localised on the opposite side from their positive-energy counterparts, resulting, as expected, in a cancellation of the skin effect whenever the products $\overline{u}u$ and $\overline{v}v$ are formed.

{\it Conclusion.---} We have constructed consistent semiclassical and second-quantised treatment of collective (elementary) excitations of a macroscopically populated continuous non-Hermitian quantum system, where the non-Hermiticity stems from a purely imaginary gauge field. The model is relevant to bosonic condensates of ultracold atoms~\cite{Tao2026,liang2022,zhao2025two} and exciton polaritons~\cite{Robin2025,Liang2026} in semiconductors, which allow for experimental probing of excitation eigenvalues and eigenstates. Most importantly, we investigated the effect of nonlinearity (stemming from inter-particle interactions in macroscopically populated condensates) on eigenstate localisation (NHSE), manifested both in the trapped condensate states and in the elementary excitation modes. The elementary excitations are treated both in the standard Bogoliubov quasiparticle framework and in the so-called hydrodynamic framework, which interprets the excitations in terms of density and phase perturbations. Remarkably, we predict that elementary excitations remain highly sensitive to the presence of non-Hermiticity and display the NHSE even when the underlying condensate states do not. This occurs, in particular, in the regime of strong nonlinearity that negates the NHSE for the trapped condensate states, but enhances the population of collective excitation modes, which still experience the effect. Our results also show that the quasiclassical and second-quantised treatments of elementary excitations in a non-Hermitian system predict different behaviour, most importantly with respect to the shifts in eigenenergies of elementary excitations. Ultimately, only experiments on the interplay between collective excitations and the non-Hermitian skin effect can resolve this discrepancy. Given that the spectroscopy of elementary excitations is well developed both for ultracold atomic~\cite{vale2021,garratt2019,vogels2002,steinhauer2002,stamper1999,lopes2017b} and exciton-polariton~\cite{Pieczarka2020,Claude2023,Claude2022,estrecho2021,stepanov2019,pieczarka2022,Bieganska2021} condensates, our theory offers an opportunity to test for quantum effects of non-Hermiticity in an experimentally accessible system.
\\

{\color{blue}{\it Acknowledgements.---}}
C. Panu is supported by the International Research Training Group (IRTG) 2675 ``Meta-ACTIVE'', Project No. 437527638. The simulations were performed on the HPC-cluster supported by EFRE Programm and Freistaat Thüringen, projects EU-0V/2020-59 (2019 FGI0017) and EU-0V/2023-1 (2022 FGI 0004). Thanks to Thomas Lettau for streamlining the simulations. E.E. is supported by the Australian Research Council (ARC) through the Discovery Early Career Researcher Award DE220100712.


\bibliographystyle{apsrev4-1}
\bibliography{Bibliografia}

\clearpage 
\newpage

\setcounter{section}{0}
\setcounter{equation}{0}%
\setcounter{figure}{0}%
\setcounter{table}{0}%

\setcounter{page}{1}

\renewcommand{\thetable}{S\arabic{table}}
\renewcommand{\theequation}{S\arabic{equation}}
\renewcommand{\thefigure}{S\arabic{figure}}
\renewcommand{\bibnumfmt}[1]{[S#1]}
\renewcommand{\citenumfont}[1]{S#1}

\onecolumngrid

\begin{center}
\textbf{\Large Supplemental Material for:\\ ``Bogoliubov excitations of a Bose-Einstein condensate in the presence of the non-Hermitian skin effect''}
\bigskip
 
\end{center}

\onecolumngrid

\appendix

\label{supplementary}

\section{Non-Hermitian quantum field theory}
We start from the following non-relativistic bosonic action: 
\begin{equation}
    \mathrm{S} = \int dt \, dx \bigg\{ \overline{\Phi} \bigg[i\partial_t -V(x) + \frac{1}{2}(\partial_x-\beta)^2 \bigg]\Phi - \frac{g}{2}\big(\overline{\Phi} \Phi\big)^2 \bigg\} 
\end{equation}
and assume the existence of a saddle point
\begin{equation}
    \frac{\delta \mathrm{S}}{\delta \overline{\Phi}} = 0 \, , \quad \frac{\delta \mathrm{S}}{\delta \Phi} = 0
\end{equation}
which results in the following equations of motion:
\begin{equation}
\begin{aligned}
       i \partial_t \Phi &= -\frac{1}{2} (\partial_x -\beta)^2 \Phi + V(x)\Phi + g \rho \Phi \\ 
       -i \partial_t \overline{\Phi} &= -\frac{1}{2} (\partial_x +\beta)^2 \overline{\Phi} + V(x)\overline{\Phi} + g \rho \overline{\Phi}
\end{aligned}
\end{equation}
where $\rho = \overline{\Phi} \Phi$. We notice that, in this context, we relax the condition of the scalar fields being complex conjugate of each other in order to fulfil the saddle point requirement. For this reason, $\overline{\Phi} \neq \Phi^*$ in general. The action maintains a global $\mathrm{U}(1)$ symmetry, \textit{i.e.} the action remains invariant under the following transformation:
\begin{equation}
    \Phi \rightarrow e^{i\alpha} \Phi \, , \quad \overline{\Phi} \rightarrow e^{-i\alpha} \overline{\Phi} \, .
\end{equation}
Noether's theorem then implies a local conservation law, \textit{i.e} considering a local $\alpha = \alpha(x, t)$, the action is modified in the following way
\begin{equation}
    \delta \mathrm{S} = \int dt \, dx \, \alpha(x, t) \bigg\{ \partial_t (\overline{\Phi} \Phi) - \frac{i}{2} \partial_x \bigg[(\partial_x \overline{\Phi})\Phi - \overline{\Phi} (\partial_x\Phi) -2\beta \overline{\Phi} \Phi \bigg] \bigg\} \overset{!}{=} 0 \, ,
\end{equation}
leading to a continuity equation
\begin{equation}
    \partial_t \rho + \partial_x j = 0
\end{equation}
with a modified Noether current with respect to the Hermitian case
\begin{equation}
    j = \frac{1}{2i} \Big[(\partial_x \overline{\Phi})\Phi - \overline{\Phi} (\partial_x\Phi) -2\beta \rho \Big] \, .
\end{equation} 
We now seek stationary solutions to the equations of motion of the form 
\begin{equation}
    \Phi(x,t) = \hat{\Phi}(x) e^{-i\mu t} \, , \quad \overline{\Phi}(x,t) = \overline{\Phi}(x) e^{i\mu t}.
\end{equation}
Substituting this form of solutions into the action leads to the following functional
\begin{equation}
    \mathcal{K} = \mathcal{H} - \mu \int dx \overline{\Phi} \Phi,
\end{equation}
where
\begin{equation}
        \mathcal{H} = \int dx \overline{\Phi}(x) \, \hat{H} \, \Phi(x) + \frac{g}{2}\int dx \, \overline{\Phi}(x) \overline{\Phi}(x) \Phi(x) \Phi(x) \, ,
\label{eq:quantum_Ham}
\end{equation}
and we defined $\hat{H} = - \frac{1}{2}(\partial_x-\beta)^2 + V(x)$. 

We are interested in the excitations (quantum fluctuations) of the classical stationary state (i.e. a condensate):  
\begin{equation}
        \hat{\Phi} =  \Phi_s + \delta\hat{\Phi} \, , \quad 
        \hat{\overline{\Phi}} = \overline{\Phi}_s + \delta\hat{{\Phi}}^{\dagger} 
\end{equation}
where, through the canonical quantisation, we promoted the fluctuation fields to operators $\overline{\Phi} \rightarrow \hat{\overline{\Phi}}$ and  $\Phi \rightarrow \hat{\Phi}$ following commutation relations
\begin{equation}
        \begin{aligned}
        \Big[\delta \hat{\Phi}(x), \ \delta\hat{\Phi}(x', t)\Big] =& \Big[ \delta\hat{\overline{\Phi}}(x, t), \ \delta\hat{\overline{\Phi}}(x', t)\Big] = 0\\
        \Big[\delta\hat{\Phi}(x, t), \ \delta\hat{\overline{\Phi}}(x', t)\Big] =& - \Big[\delta\hat{\overline{\Phi}}(x, t), \ \delta\hat{\Phi}(x', t)\Big] = \delta(x-x') \, .
    \end{aligned}
    \label{eq:fluctuations}
\end{equation}
The commutation relations are not bosonic since $\delta\hat{\overline{\Phi}} \neq \delta\hat{\overline{\Phi}}$. Note that here, we are considering the condensate described by a fully classical function, but one can also construct a quantum framework considering $\hat{\Phi} =  \Phi_0 \, \hat{a}_0 + \delta\hat{\Phi}$ and $\big[\delta \hat{\Phi} (x), \, \delta \hat{\overline{\Phi}}(x')\big] = \delta(x-x') - \Phi_0(x) \overline{\Phi}_0(x')$. 

We substitute Eqs. (\ref{eq:fluctuations}) into Eq. (\ref{eq:quantum_Ham}) obtaining
\begin{equation}
\begin{aligned}
        \hat{\mathcal{H}} =& \int dx \Big(\Phi_s + \delta\hat{\Phi}\Big) \, \hat{H} \, \Big(\overline{\Phi}_s + \delta\hat{{\Phi}}^{\dagger}\Big)
         + \frac{g}{2}\int dx \, \Big(\overline{\Phi}_s + \delta\hat{{\Phi}}^{\dagger}\Big) \Big(\overline{\Phi}_s + \delta\hat{{\Phi}}^{\dagger}\Big) \Big(\Phi_s + \delta\hat{\Phi}\Big) \Big(\Phi_s + \delta\hat{\Phi}\Big) \, = \\
         =& \int dx \bigg[\delta\hat{\overline{\Phi}} \hat{H} \delta \hat{\Phi} + 2 g \rho_s \delta \hat{\overline{\Phi}} \delta \hat{\Phi} + \frac{g}{2} \Big(\overline{\Phi}_s\Big)^2 \delta \hat{\Phi}^2 + \frac{g}{2} \Big(\Phi_s\Big)^2 \delta \hat{\overline{\Phi}}^2 + \\
         & \hspace{3 cm} + \overline{\Phi}_s \Big(\hat{H} + \frac{g}{2} \rho_s \Big)\Phi_s + \delta \hat{\overline{\Phi}}\Big(\hat{H} + g \rho_s \Big)\Phi_s + \overline{\Phi}_s \Big(\hat{H} + g \rho_s \Big) \delta \hat{\Phi} \bigg] \, ,
\end{aligned}
\label{eq:ham_sub}
\end{equation}
and 
\begin{equation}
    \hat{\mathcal{K}} = \hat{\mathcal{H}} - \mu \int dx \hat{\overline{\Phi}} \hat{\Phi} =  \hat{\mathcal{H}} - \mu \int dx\Big(\rho_s + \overline{\Phi}_s \delta \hat{\Phi} + \Phi_s \delta \hat{\overline{\Phi}} + \delta \hat{\Phi} \delta\hat{\overline{\Phi}}\Big) \, ,
\end{equation}
where $\rho_s = \overline{\Phi}_s \Phi_s$. We can then require $\Phi_s$ and $\overline{\Phi}_s$ to be solutions of the Gross-Pitaevskii equations: 
\begin{equation}
\begin{aligned}
    \big(&\hat{H} + g \rho_s - \mu \big) \Phi_s = 0 \\
    \big(&\hat{H}^T + g \rho_s - \mu \big) \overline{\Phi}_s = 0 \, , 
\end{aligned}
\end{equation}
where $\hat{H}^T$ is the formal transpose of $\hat{H}$, defined as 
\begin{equation}
    \int dx \Phi \hat{H} \phi = \int dx \phi \hat{H}^T \Phi \, .
\end{equation}
For example, if $\hat{H} = (\hat{p} + A)^2 + V$, then $\hat{H}^T = (\hat{p} - A)^2 + V$ and comes from the partial integration of the corresponding term in Eq. (\ref{eq:ham_sub})
\begin{equation}
    \int dx \overline{\Phi}_s \Big(\hat{H} + g \rho_0 \Big) \delta \hat{\Phi} = \int dx  \delta \hat{\Phi} \Big(\hat{H}^T + g \rho_s \Big) \overline{\Phi}_s
\end{equation}
If we discard the kinetic term in the Thomas-Fermi limit, we find 
\begin{equation}
    \rho_s(x) \equiv  \overline{ \Phi}_s(x) \Phi_s(x) = \frac{\tilde{\mu} - V(x)}{g} \, \theta({\tilde{\mu} - V(x)})\, .
\end{equation}
where $\tilde{\mu} = \mu + A^2$.
There is a residual functional degeneracy leaving $\rho_s$ invariant, $\Phi_s(x) \rightarrow \lambda(x)\Phi_s(x)$ and $ \overline{ \Phi}_s(x) = \overline{ \Phi}_s(x) / \lambda(x)$, that is only lifted from the kinetic term. For example in the presence of an imaginary gauge field $A=i\beta$, the density $\rho_s$ is left invariant and $\Phi_s(x) \rightarrow \lambda(x)\Tilde{\Phi}_s(x)$ and $ \overline{ \Phi}_s(x) = \Tilde{\Phi}^*_s(x) / \lambda(x)$, with  $\lambda(x)=e^
{-\beta x}$.

Going back to the general problem, we obtain: 
\begin{equation}
        \hat{\mathcal{K}} = E_s \big( \Phi_s, \overline{\Phi}_s\big) + \int dx \, \delta \hat{\overline{\Phi}} \big(\hat{H} - \mu + 2g \rho_s\big) \delta \hat{\Phi} + \int dx \, \frac{g}{2} \Big[\big(\overline{\Phi}_s \big)^2 \delta \hat{\Phi}^2 + \Phi_s^2 \big(\delta \hat{\overline{\Phi}}\big)^2\Big] \, , 
\end{equation}
where we defined the energy functional
\begin{equation}
    E_s \big( \Phi_s, \overline{\Phi}_s\big) = \int dx \,\overline{\Phi}_s \Big(\hat{H} + \frac{g}{2} \rho_s  \Big)\Phi_s \, .
\end{equation}
To diagonalise $\hat{\mathcal{K}}$, we perform a Bogoliubov transformation by defining the following creation and annihilation operators 
\begin{equation}
\begin{aligned}
    \hat{\bar{\gamma_i}} =& \int dx \big[u_i(x) \delta \hat{\overline{\Phi}}(x) + v_i(x)  \delta \hat{\Phi}(x) \big] \\
    \hat{\gamma_i} =&  \int dx \big[\bar{u}_i(x)  \delta \hat{\Phi}(x) + \bar{v}_i(x) \delta \hat{\overline{\Phi}}(x) \big]
\end{aligned}
\qquad \mathrm{with} \qquad 
\begin{aligned}
    \big[&\hat{\gamma}_i, \, \hat{\gamma}_j\big] = \big[\hat{\bar{\gamma}}_i, \, \hat{\bar{\gamma}}_j\big] = 0 \\
    \big[&\hat{\gamma}_i, \, \hat{\bar{\gamma}}_j\big] = -\big[\hat{\bar{\gamma}}_i, \, \hat{\gamma}_j\big] = \delta_{ij}
\end{aligned}
\label{eq:bdg_operator}
\end{equation}
In order to satisfy the commutation rules, one needs the following orthogonality relations
\begin{equation}
    \int dx \big(\bar{u}_i u_j - \bar{v}_i v_j\big) = \delta_{ij} \quad \mathrm{and} \quad  \int dx \big(u_i v_j - v_i u_j\big) = \int dx \big(\bar{u}_i \bar{v}_j - \bar{v}_i \bar{u}_j\big) = 0 \, .
\end{equation}
The diagonalisation procedure allows us to write the $\hat{\mathcal{K}}$ operator in the form
\begin{equation}
    \hat{\mathcal{K}}_{\mathrm{D}} = E_s + \sum_i \epsilon_i \hat{\bar{g}}_i \hat{g}_i
\end{equation}
that satisfies two sets of commutation conditions 
\begin{equation}
    \big[\hat{\bar{\gamma}}_i, \, \hat{\mathcal{K}}_{\mathrm{D}}\big] = -\epsilon_i \hat{\bar{\gamma}}_i \qquad \mathrm{and} \qquad \big[\hat{\gamma}_i, \, \hat{\mathcal{K}}_{\mathrm{D}}\big] = \epsilon_i \hat{\gamma}_i \, ,
    \label{eq:bdg_commutators}
\end{equation}
where we used the identity 
\begin{equation}
    [a, bc] = [a, b]c + b[a, c] \, .
    \label{eq:leibniz_rule}
\end{equation}
Substituting Eq. (\ref{eq:bdg_operator}) into (\ref{eq:bdg_commutators}),  using the identity (\ref{eq:leibniz_rule}) together with 
\begin{equation}
    \int dx' f(x') \frac{d \delta(x-x')}{dx} = - \frac{df(x)}{dx},
\end{equation}
and collecting the terms with $\delta \hat{\Phi}$ and $\delta$, we obtain the following set of equations  $\hat{\overline{\Phi}}$
\begin{equation}
    \begin{cases}
        (\hat{H} + 2g\rho_s - \mu)u - g \big(\Phi_s\big)^2 v = \epsilon u \\
        (-\hat{H}^T - 2g\rho_s + \mu) v + g \big(\overline{\Phi}_s\big)^2 u = \epsilon v
    \end{cases}
\end{equation}
together with the corresponding solutions for $\epsilon \rightarrow - \epsilon$,  $u \rightarrow \bar{v}$ and $v \rightarrow \bar{u}$
\begin{equation}
    \begin{cases}
        (-\hat{H}^T - 2g\rho_s +\mu) \bar{u} + g \big(\overline{\Phi}_s\big)^2 \bar{v} = -\epsilon \bar{u} \\
        (\hat{H} + 2g\rho_s -\mu) \bar{v} - g \big(\Phi_s\big)^2 \bar{u} = -\epsilon \bar{v} \, .
    \end{cases}
\end{equation}
The non-Hermitian generalisation of the Bogoliubov-de Gennes operator takes the form: 
\begin{equation}
        \hat{\mathcal{L}} = 
         \begin{pmatrix}
        \hat{H} -\mu + 2g \rho_s & -g \Phi_s^2 \\
        g(\overline{\Phi}_s)^2 & -\hat{H}^T + \mu - 2g \rho_s 
    \end{pmatrix}
\end{equation}
and the excitation operators are written as:
\begin{equation}
   \delta \hat{\Phi}(x) = \sum_i u_i(x) \hat{\gamma}_i - \bar{v}_i(x) \hat{\bar{\gamma}}_i \quad \mathrm{and} \quad \delta \hat{\overline{\Phi}}(x) = \sum_i \bar{u}_i(x) \hat{\bar{\gamma}}_i - v_i(x) \hat{\gamma}_i \, .
\end{equation}
Taking the ensemble average $\langle ... \rangle$ of the density operator we obtain
\begin{equation}
\begin{aligned}
    \langle\delta \hat{\overline{\Phi}} \delta \hat{\Phi}\rangle =& \langle \sum_{ij} \big(\bar{u}_i \hat{\bar{\gamma}}_i - v_i \hat{\gamma}_i\big) \big(u_j \hat{\gamma}_j - \bar{v}_j \hat{\bar{\gamma}}_j\big) \rangle = \\
    =& \sum_i \bar{u}_i u_i \langle \hat{\bar{\gamma}}_i \hat{\gamma}_i \rangle + \bar{v}_i v_i \langle \hat{\bar{\gamma}}_i \hat{\gamma}_i \rangle - \bar{v}_i v_i
\end{aligned}
\end{equation}

\section{Modelling the skin effect in the continuum.}
\label{sec:model}
The non-Hermitian skin effect can be described in the framework of a continuum model with minimal coupling by including an imaginary vector potential. To understand the reason behind this choice, we start by reviewing the model for a quantum particle in the presence of electromagnetic fields
\begin{equation}
    \mathbf{E} = -\mathbf{\nabla}\phi - \frac{1}{c} \frac{\partial \mathbf{A}}{\partial t} \, , \quad \mathbf{B = \mathbf{\nabla}\times \mathbf{A}} \, .
\end{equation}
where $\phi$ is a scalar electric potential and $\mathbf{A}$ is a vector magnetic potential.

We consider the following Hamiltonian describing a confined particle with a charge $q$:
\begin{equation}
    H = \frac{\big(\mathbf{p}-\frac{q}{c}\mathbf{A}\big)^2}{2m} + V(\mathbf{r}) + q\phi.
    \label{eq:H_harm_osc}
\end{equation}
 To write the Hamiltonian in this form, we used the Hamiltonian without the electromagnetic fields, $H_0$, and applied the \textit{minimal coupling} approach by making the substitutions: 
\begin{equation}
    \mathbf{p} \rightarrow \mathbf{p} - \frac{q}{c}\mathbf{A} \, , \quad H_0 \rightarrow H_0 + q \, \phi
\end{equation}
The Hamiltonian is gauge invariant, \textit{i.e.} after the transformation
\begin{equation}
    \mathbf{A} \rightarrow\mathbf{A} + \mathbf{\nabla}\Lambda \, , \quad \phi \rightarrow \phi - \frac{1}{c}\frac{\partial \Lambda}{\partial t} 
\end{equation}
the resulting new Hamiltonian $H'$ will describe the same time evolution of $H$. The two time evolutions are simply linked by a phase transformation, so the constant term $+q\phi$ (a shift in the spectrum) is not relevant for the dynamics and will be discarded from now on. 

The gauge field has an interesting effect on the phase of the wavefunction. It can be proven that, after performing a gauge transformation, one has
\begin{equation}
    \Psi' = e^{iF}\psi \, , \quad \text{where} \quad F= \frac{q}{c}\Lambda(\mathbf{r}) \, 
\end{equation}
A widely used approach to introducing the minimal coupling in a discrete (lattice) model on lattice for Eq. (\ref{eq:H_harm_osc}) is called \textit{Peierls substitution}. For simplicity, we consider a 1D chain, but the result can be easily generalised to higher dimensions. In the tight-binding framework, the second order derivative corresponds to
\begin{equation}
    \partial^2_x \rightarrow \frac{1}{a^2} \sum_i \big(\ketbra{i+1}{i} + \ketbra{i}{i+1} -2\ketbra{i}{i} \big),
\end{equation}
where $\ket{i} \equiv \ket{ai} = \ket{x}$ represent the discretised positional states and $\ket{\psi(x)} = \sum_ia_i \ket{i}$. Then our Hamiltonian becomes
\begin{equation}
    \hat{H} = \sum_{i} \big(V(ai) -2t\big) \, \ketbra{i}{i} -t  \sum_{i} \big( e^{i\phi_i} \, \ketbra{i+1}{i} + e^{-i\phi_i} \, \ketbra{i}{i+1} \big) \, 
\end{equation}
where $a$ is the distance between the sites of the lattice, $t\equiv \hbar^2/(2ma^2)$ and $\hat{c}^{\dagger}_i$ the particle creation operator on site $i$ 
\begin{equation}
    \phi_i = \frac{2\pi q}{h} \int^{(i+1)a}_{ia} A(x) \, dx \simeq \frac{q a}{\hbar c} A(ai)
\end{equation}
assuming $A(x)$ slowly varies between lattice sites. The translation operator can be written in terms of the momentum operator as 
\begin{equation}
    \ketbra{i+1}{i} = e^{-i\hat{p}_x a/\hbar}\ketbra{i}{i} = \big(1-\frac{i\hat{p}_x a}{\hbar} -\frac{\hat{p}^2_x a^2}{2\hbar} + \mathcal{O}(a^3)\big)
\end{equation}
where we performed the expansion assuming small $a$, and similarly 
\begin{equation}
    e^{i\phi} = 1+ i\phi - \frac{1}{2}\phi^2 + \mathcal{O}(a^3) \, , \quad \mathrm{with} \quad \phi \simeq \frac{qa}{\hbar c} A \, .
\end{equation}
Substituting the two expansions, we obtain
\begin{equation}
   e^{i\phi_i} \ketbra{i+1}{i} +  e^{-i\phi_i} \ketbra{i}{i+1} = \Big(-\frac{a^2}{\hbar^2}\big(p_x - \frac{q}{c} A(ia)\big)^2 + 2 + \mathcal{O}(a^3)\Big) \ketbra{i}{i} \, .
\end{equation}
In this way, in the continuum limit we again obtain the Hamiltonian in the familiar form:
\begin{equation}
    \hat{H} = \frac{\big(p-\frac{q}{c}A\big)^2}{2m} + V(x) \, .
\end{equation}

Here, our purpose is to construct a continuous model for the skin effect, which is typically studied in discrete models (lattices) with nonreciprocal inter-site hoppings. The simplest way to have asymmetrical hopping it is to consider a complex vector potential with the magnitude $A =iA_0$. In this way the exponentials in the Peierls subtstitution will not only become a simple phase, but will give different weights to the hoppings: 
\begin{equation}
    t_{\pm}= -te^{\pm \frac{qa}{\hbar c}A_0} \, .
\end{equation}
This results in the so-called Hatano-Nelson (HN) model:
\begin{equation}
    \hat{H_i} = \epsilon_i \, \ketbra{i}{i} + t_- \, \ketbra{i+1}{i} + t_+ \, \ketbra{i}{i+1} \, ,
\end{equation}
which is the simplest model featuring the non-Hermitian skin effect. 

The continuum analogy of the HN model is then described by the Hamiltonian:
\begin{equation}
\hat{H}_{\mathrm{HN}}= \frac{\big(p-i\tilde{\beta}\big)^2}{2m} + V(x) \, ,
\end{equation}
where $\tilde{\beta} \equiv qA_0/c$ because the minimal coupling now has a different physical meaning.

Considering a particular form of the harmionic trapping potential $V(x)~=~m\omega^2x^2/2$, the eigenstates of the Hamiltonian will be modified in the following way:
\begin{equation}
    \psi'_n = e^{-A_0 x} \psi_n(x) \, = \frac{1}{(l^2\pi)^{1/4}\sqrt{2^n n!}}\mathcal{H}_n(x/l) e^{-((x/l\sqrt{2})^2 + \tilde{\beta} x)}
    \label{eq:HO_spectra}
\end{equation}
where $l=\sqrt{\hbar/m\omega}$ and $\mathcal{H}_n$ are the Hermite polynomials, defined as following
\begin{equation}
    \mathcal{H}_n(\xi) = e^{\xi^2/2}\Big(\xi - \frac{d}{d\xi}\Big)^n e^{-\xi^2/2} \, .
\end{equation}
The corresponding eigenvalues are:
\begin{equation}
    E_n = \hbar \omega(1/2 + n) \, \quad \mathrm{with} \quad n \in \mathbb{N} \, .
\end{equation}
Introducing a dimensionless variable $\xi = x/l$, we obtain:
\begin{equation}
    \hat{H}_{\mathrm{HN}} \psi =-\frac{1}{2}\Big(\partial_{\xi} + \beta\Big)^2\psi + \frac{1}{2}\xi^2 \psi = \frac{E}{\hbar \omega} \psi \, ,
\end{equation}
where $\beta = qlA_0/c$.

In the context of the Gross-Pitaevskii Eq. \ref{eq:gross_pita}, the dimensionless units need to take into account the normalisation $\int |\psi| dx = N$. In our treatment, we rescale the function $\psi \equiv \sqrt{N/l} \, \Phi $ such as $\int |\Phi|^2 d \xi = 1$. In the context of a particular experimental setting, one can retrieve the interaction strength using the following relation: 
\begin{equation}
    g = \frac{g_e N}{\hbar \omega l} \, . 
\end{equation}
\newpage

\subsection{BdG excitations for a harmonic potential}
The aim of this section is to compare the solutions to the BdG problem (\ref{eq:bdg_problem}) obtained in the main text with those obtained the Hermitian limit. Following the procedure discussed in \cite{Gao2020}, we present the deviation of the dipole mode excitation, \textit{i.e.} the state oscillating with the harmonic trap frequency. 

The ground state of the GPE equation (\ref{eq:gross_pita}) is unique up to a phase factor $\Phi_0 = e^{i\theta} |\Phi_0|$ and this affects the BdG eigenstates as a simple phase shift $(u, \ v)^T \rightarrow (e^{i\theta}u, \ e^{-i\theta}v)^T$. For this reason, we can choose $\Phi_0$ to be real and positive. We start from the GPE for a stationary solution:
\begin{equation}
    H \Phi + g \, \Phi^2 \, \Phi = \mu \Phi \, , \quad \mathrm{where} \quad H =  -\big(\partial_x + \beta\big)^2 /2 + x^2/2 \, .
\end{equation}
We then apply the operator $\partial_x+\beta$ to both sides of the GPE equation and multiply by $x$, which results in two new equations:
\begin{equation}
\begin{aligned}
    \Big[&H + 3g \Phi^2 -\mu\Big](-\partial_x -\beta) \Phi + 2 \beta g \Phi^2 = x \Phi \\
    \Big[&H + g \Phi^2 -\mu\Big] x \Phi = (-\partial_x - \beta)\Phi \, .
\end{aligned}
\label{eq:gpe_modi}
\end{equation}
On the other hand, if we perform the following change of basis
\begin{equation}
    \begin{pmatrix}
        \tilde{u} \\
        \tilde{v}
    \end{pmatrix} = \frac{1}{\sqrt{2}}
    \begin{pmatrix}
        1 & 1 \\
        1 & -1
    \end{pmatrix}
    \begin{pmatrix}
        u \\
        v
    \end{pmatrix} = M
     \begin{pmatrix}
        u \\
        v
    \end{pmatrix}
    \quad \rightarrow \quad 
    \tilde{\mathcal{L}} = M^{-1} \mathcal{L}M = 
    \begin{pmatrix}
        0 & H - \mu + 3g\Phi_0^2 \\
        H + \mu + g \Phi_0^2 & 0
    \end{pmatrix}  \equiv
    \begin{pmatrix}
        0 & H_+ \\
        H_- & 0
    \end{pmatrix} \, ,
\end{equation}
with $H$ being real, we notice that the Eqs. (\ref{eq:gpe_modi}) solve the eigenvalue problem for $\tilde{\mathcal{L}}$ in the limit of $\beta\rightarrow 0$. The eigenvalue is the harmonic trapping frequency ($\omega=1$ in our units), while the eigenvectors are 
\begin{equation}
\begin{pmatrix}
    \tilde{u} \\
    \tilde{v}
\end{pmatrix} = 
\begin{pmatrix}
    x \Phi \\
    -\partial_x \Phi
\end{pmatrix} \quad  \rightarrow \quad 
\begin{pmatrix}
    u \\
    v
\end{pmatrix} = M^{-1}
\begin{pmatrix}
    \tilde{u} \\
    \tilde{v}
\end{pmatrix} = \frac{1}{\sqrt{2}}
\begin{pmatrix}
    x \Phi - \partial_x \Phi \\
    x \Phi + \partial_x \Phi
\end{pmatrix}
\end{equation}
After this rotation we obtain a new normalisation condition $\mathrm{Re} \int \tilde{u} \tilde{v} \, dx = 1/2$, which can be easily proven for the obtained solutions. 

The solution above is a special case; we can now analyse the more general Thomas-Fermi limit. For $g\gg 1$ we can discard the kinetic terms in the GPE, and the solution becomes: 
\begin{equation}
    \Phi_{\small \mathrm{TF}} = \frac{\mu - V(x) + \beta^2/2}{g} \theta(\mu - V(x)) \qquad \mathrm{and} \qquad \mu = \frac{1}{2}\bigg(\frac{3g}{2}\bigg)^{2/3}
\end{equation}
where $V(x)=x^2/2$, $\theta(x)$ is the Heaviside function, and $\mu$ is obtained by the normalisation condition. Since $H_+$ is positive definite, we can write
\begin{equation}
\tilde{v} = H_+^{-1} \epsilon \tilde{u} \quad \rightarrow \quad H_+ H_- \tilde{u} = \epsilon^2 \tilde{u}
\end{equation}
and substituting $\Phi_{\small \mathrm{TF}}$ we obtain
\begin{equation}
    \Big[ -\frac{1}{2}(\partial^2_x+2\beta\partial_x) + 2\mu - x^2\Big] \Big[ -\frac{1}{2}(\partial^2_x+2\beta\partial_x) \Big]\tilde{u} = \epsilon^2 \tilde{u} \, .
\end{equation}
In the limit $\mu \gg 1$ ($g\gg 1$), we have
\begin{equation}
    (x^2 - 2\mu)(\partial_x+\beta)^2 \tilde{u} - \beta^2 (x^2-2\mu) \tilde{u}= 2\epsilon^2 \tilde{u}
\end{equation} 
With the following rescaling $\tilde{x} = x/\sqrt{2\mu}$, the previous equation becomes: 
\begin{equation}
    (\tilde{x}^2-1)(\partial_{\tilde{x}} - \sqrt{2\mu} \beta)^2 \tilde{u}(\tilde{x}) + 2\mu \beta^2 (\tilde{x}^2-1) \tilde{u}(\tilde{x})= 2 \epsilon^2 \tilde{u}(\tilde{x}) \, .
\end{equation}
We can then seek solutions of the form $\tilde{u}(\tilde{x}) = \mathrm{exp}(-\sqrt{2\mu} \beta \tilde{x}) f(\tilde{x})$ 
\begin{equation}
    (\tilde{x}^2-1)\partial_{\tilde{x}}^2 f(\tilde{x}) - 2\mu \beta^2 (\tilde{x}^2-1) f(\tilde{x}) = 2 \epsilon^2 f(\tilde{x})  
\end{equation}
Taking the limit of $\beta=0$ and taking a derivative on both sides, we obtain:
\begin{equation}
    (\tilde{x}^2-1) z''(\tilde{x}) + 2\tilde{x} z'  = 2 \epsilon^2 z \, , 
\end{equation}
where $z = f'(x)$. This is a Sturm-Liouville problem with boundary conditions $z(\tilde{x}=\pm 1) = 0$ known as Legendre equations, which is known to have the following solutions for the eigenvalues:
\begin{equation}
    2\epsilon^2 = l(l-1) \quad \rightarrow \quad \epsilon = \pm \sqrt{\frac{l(l-1)}{2}}
\end{equation}

\newpage

\subsection{Hydrodynamic equations}
We start once again from the Gross-Pitaevskii equation 
\begin{equation}
    i\partial_t \Phi  = -\frac{1}{2}(\partial_x + \beta)^2 \Phi + V(x) \Phi + g |\Phi|^2 \Phi \, ,
\end{equation}
then we multiply it by $\Phi^*$ and subtract its own complex conjugate obtaining 
\begin{equation}
    \partial_t |\Phi|^2 = -(\partial_x + 2\beta) \frac{1}{2i} \big[\Phi^* \partial_x\Phi - \Phi \partial_x \Phi^*\big].
\end{equation}
This equation can be rewritten in the more convenient form:
\begin{equation}
    \partial_t n + (\partial_x +2\beta) \, nv = 0 \, ,  \quad \mathrm{where} \quad n \equiv |\Phi|^2 \quad \mathrm{and} \quad v \equiv \frac{1}{2i} \frac{\Phi^*\partial_x \Phi - \Phi \partial_x \Phi^*}{n} \, .
\end{equation}
We now write the equations for the amplitude and phase of the wavefunction $\Phi = \sqrt{n} e^{i\varphi}$. It can be easily proven that $v = \partial_x \varphi (x)$, which allows us to obtain the following set of equations:
\begin{equation}
    \begin{aligned}
        \partial_t n =& - (\partial_x + 2\beta) \, n \partial_x \varphi  \\
        -\partial_t \varphi =& -\frac{1}{2\sqrt{n}} (\partial_x + \beta)^2 \sqrt{n} + \frac{1}{2}v^2 + V(x) + gn
    \end{aligned}
    \label{eq:hydro_eq}
\end{equation}
Taking a spatial derivative of the second equation, we obtain an expression for the velocity: 
\begin{equation}
    \partial_t v = -\partial_x \big(\tilde{\mu} + \frac{1}{2}v^2\big) \quad \mathrm{where} \quad \tilde{\mu} =  -\frac{1}{2\sqrt{n}} (\partial_x + \beta)^2 \sqrt{n} + V(x) + gn \, .
    \label{eq:vel_eq}
\end{equation}
We now linearise the equations around an equilibrium state with the density $n_{\small \mathrm{eq}}$ and a vanishing velocity $v_{\small \mathrm{eq}} = 0$. Discarding the second order terms in velocity and density perturbations for Eq. (\ref{eq:vel_eq}), we obtain: 
\begin{equation}
    \partial_t \delta v = -\partial_x \delta \mu \qquad \mathrm{with} \qquad \delta \mu = \frac{1}{4\sqrt{n_{\small \mathrm{eq}}}}\bigg[ \frac{\delta n}{n_{\small \mathrm{eq}}}(\partial^2_x + 2\beta \partial_x)\sqrt{n_{\small \mathrm{eq}}} - (\partial^2_x + 2\beta \partial_x)\Big(\frac{\delta n}{n_{\small \mathrm{eq}}} \Big)\bigg] + g \, \delta n \, .
\end{equation}
In order to obtain an equation for the density perturbation we take the time derivative of the first equation in (\ref{eq:hydro_eq}) resulting in 
\begin{equation}
    \partial^2_t \delta n = (\partial_x +2\beta) [n_{\small \mathrm{eq}} \partial_x \delta \mu] \, .
\end{equation}
In the Thomas-Fermi regime, we discard the kinetic terms of $\delta \mu$ and obtain the following modified hydrodynamic equation for the density:
\begin{equation}
    \partial_t^2 \delta  n = (\partial_x +2\beta) [gn_{\small \mathrm{TF}} \partial_x \delta n ]
    \label{eq:hydro_TF}
\end{equation}
and for the stationary solutions:
\begin{equation}
     (\partial_x +2\beta) [gn_{\small \mathrm{TF}} \partial_x \delta n ] = -\epsilon^2 \delta n
    \label{eq:hydro_TF}
\end{equation}

For completeness, we also show how to relate the BdG equations to the hydrodynamic approach. In the BdG framework, we write the perturbations of the condensate in the following way
\begin{equation}
    \Phi(x, t) = e^{-i\mu t}\bigg[\sqrt{n_{\small \mathrm{eq}}(x)} e^{i\varphi_{\small \mathrm{eq}}(x)} + \delta \Phi(x, t) \bigg] \, , \quad \mathrm{where} \quad \delta \Phi(x, t) = \sum_i \bigg[ u_i(x) e^{-i\epsilon_it} - v^*(x) e^{+i\epsilon_it} \bigg]
\end{equation}.
On the other hand, a perturbation $\delta \Phi$ can be written in terms of the density and phase perturbations, $n(x, t) = n_{\small \mathrm{eq}}(x) + \delta n(x, t)$ and $\varphi(x, t) = \varphi_{\small \mathrm{eq}}(x) + \delta \varphi(x, t)$ respectively, of the equilibrium state in the following way 
\begin{equation}
    \delta \Phi(x, t) =  \bigg[\frac{\delta \, n(x, t)}{2 n_{\small \mathrm{eq}}(x)} + i \delta \varphi (x, t)\bigg] \sqrt{n_{\small \mathrm{eq}}(x)} \, .
\end{equation}
Here, the time dependent density and phase perturbations are real quantities and can be written in terms of stationary waves as:
\begin{equation}
    \delta n(x, t) = \sum_i \bigg[ \delta n_i(x)e^{-i\epsilon_i t} + \delta n_i^*(x)e^{+i\epsilon_i t} \bigg]
\end{equation}
and similarly for $\delta \varphi(x, t)$. Matching the exponential terms in the two expressions for $\delta \Phi$, we obtain the relations: 
\begin{equation}
\begin{aligned}
       \delta n_i (x) =& \sqrt{n_{\small \mathrm{eq}}(x)} \bigg[u_i(x)e^{-i\varphi_{\small \mathrm{eq}}(x)}  - v_i(x)e^{+i\varphi_{\small \mathrm{eq}}(x)}\bigg] \\
       \delta \varphi_i (x) =& \frac{1}{2i\sqrt{n_{\small \mathrm{eq}}(x)}} \bigg[u_i(x)e^{-i\varphi_{\small \mathrm{eq}}(x)}  + v_i(x)e^{+i\varphi_{\small \mathrm{eq}}(x)}\bigg] \, .
\end{aligned}
\end{equation}
We point out that $\delta n(x)$ and $\delta \varphi(x)$ are complex quantities and can result in a dephasing from the equilibrium state. In fact if $\delta n (x) = |\delta n(x)| \, \mathrm{exp}(i \phi_n(x))$ then $\delta n (x, t) = |\delta n(x)| \mathrm{cos}\big(\epsilon t -\phi_n (x)\big)$ and in a moving perturbation with velocity $v(x, t) = \partial_x \delta \varphi(x, t) = |\delta \varphi'(x)| \cos\big(\epsilon t - \phi_{\delta \varphi'}(x)\big)$ as already discussed.

In the case of a real equilibrium state (or a state with a constant phase), from Eq. \ref{eq:bdg_operator_mt} we find that the BdG equations can be rewritten as 
\begin{equation}
\begin{aligned}
    \bigg[-\frac{1}{2}(\partial_x + \beta)^2 + V(x) - \mu + 3g n_{\small \mathrm{eq}} \bigg]\frac{\delta  n}{\sqrt{n_{\small \mathrm{eq}}}} = \epsilon \,   2 i \sqrt{n_{\small \mathrm{eq}}} \delta \varphi \\
    \bigg[-\frac{1}{2}(\partial_x + \beta)^2 + V(x) - \mu + gn_{\small \mathrm{eq}} \bigg] 2 i \sqrt{n_{\small \mathrm{eq}}} \delta \varphi = \epsilon \, \frac{\delta  n}{\sqrt{n_{\small \mathrm{eq}}}}.
\end{aligned}
\end{equation}
Within the Thomas-Fermi approximation, we then obtain:
\begin{equation}
\begin{aligned}
    2g \delta  n = \epsilon \,   2 i \delta \varphi \\
    \sqrt{n_{\small \mathrm{TF}}} \, (\partial^2_x + \beta \partial_x) [\sqrt{n_{\small \mathrm{TF}}} \delta \varphi] = i \epsilon \, \delta  n \, .
\end{aligned}
\end{equation}
where $n_{\small \mathrm{TF}} = (\mu - V(x) + \beta^2/2)/g$.
In this regime, we discard any term assuming the equilibrium kinetic term relative to its envelope is much smaller compared to the one of the perturbation one
\begin{equation}
    \bigg| \frac{\partial_x n_{\small \mathrm{TF}}}{n_{\small \mathrm{TF}}}\bigg| \ll \bigg| \frac{\partial_x\delta n}{\delta n} \bigg|.
\end{equation}
Then we can simplify the second equation and rewrite it in terms of $\delta n$ 
\begin{equation}
    (\partial_x + 2\beta) [g n_{\small \mathrm{TF}} \, \delta n] = -\epsilon^2 \delta n
\end{equation}
as obtained with the hydrodynamic approach.
\begin{figure}[H]
    \centering
    \includegraphics[width=0.6\linewidth]{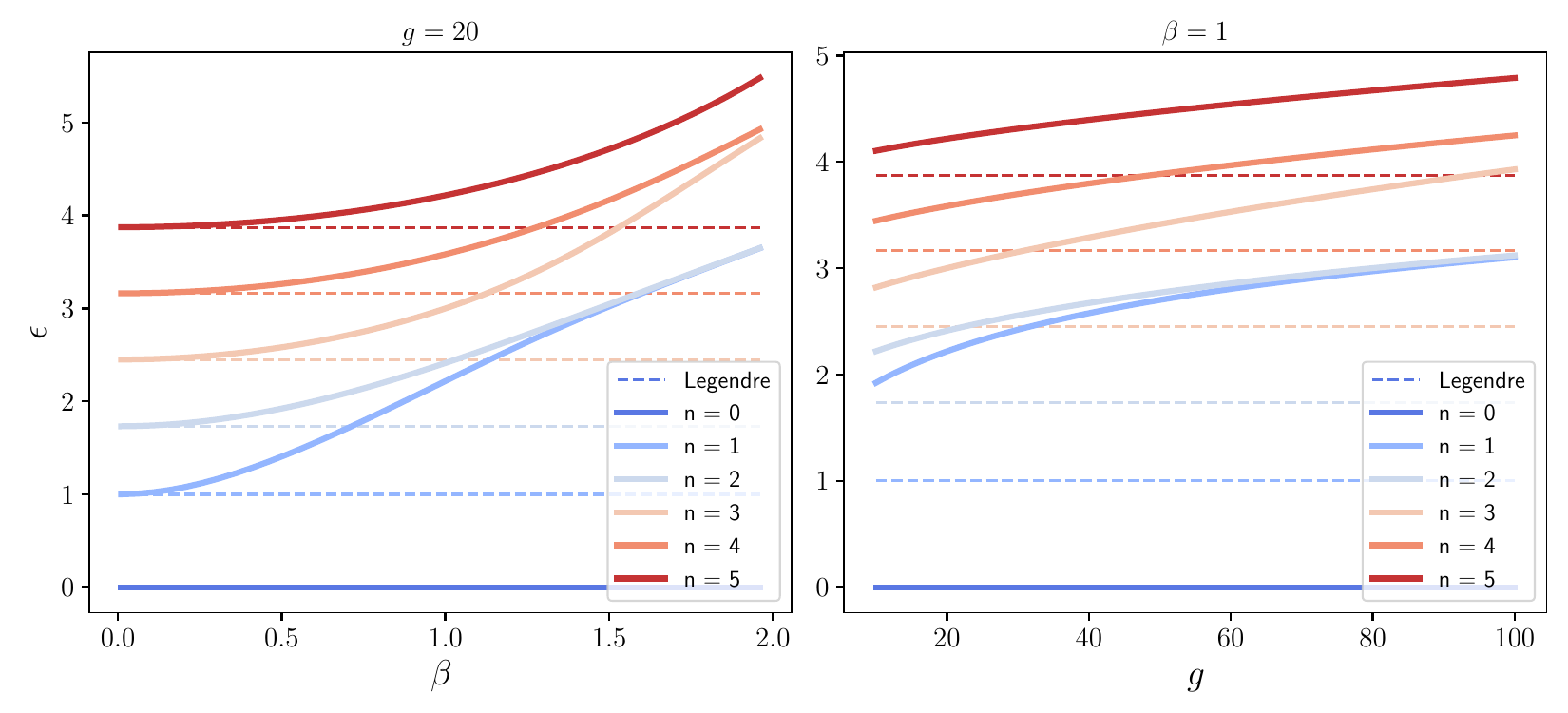}
    \caption{Eigenvalues found by solving Eq.\ref{eq:hydro_eq} using Wolfram Mathematica with varying $\beta$ and $g$ on the left and right panels, respectively.}
    \label{fig:mathematica_eigenvalues}
\end{figure}

\subsection{Supergaussian trapping potential}
\begin{figure}[H]
\centering
    \begin{subfigure}[b]{0.3\linewidth}
        \centering
        \begin{overpic}[width=\linewidth]{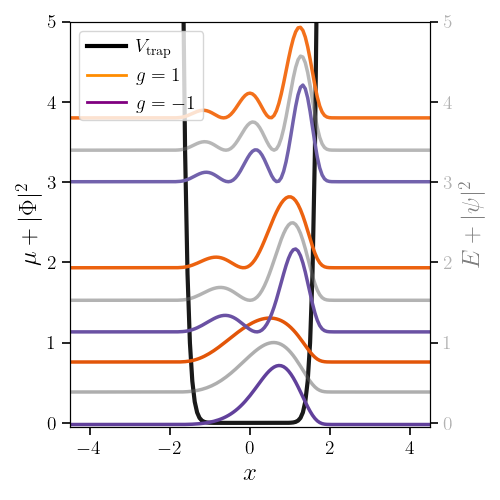}
        \put(-3,142){(a)}
        \end{overpic}
        \phantomcaption
        \label{fig:nonlinear_combined_a}
    \end{subfigure}
     \begin{minipage}[b]{0.3\linewidth}
    \begin{subfigure}[b]{\linewidth}
        \centering
        \begin{overpic}[width=\linewidth]{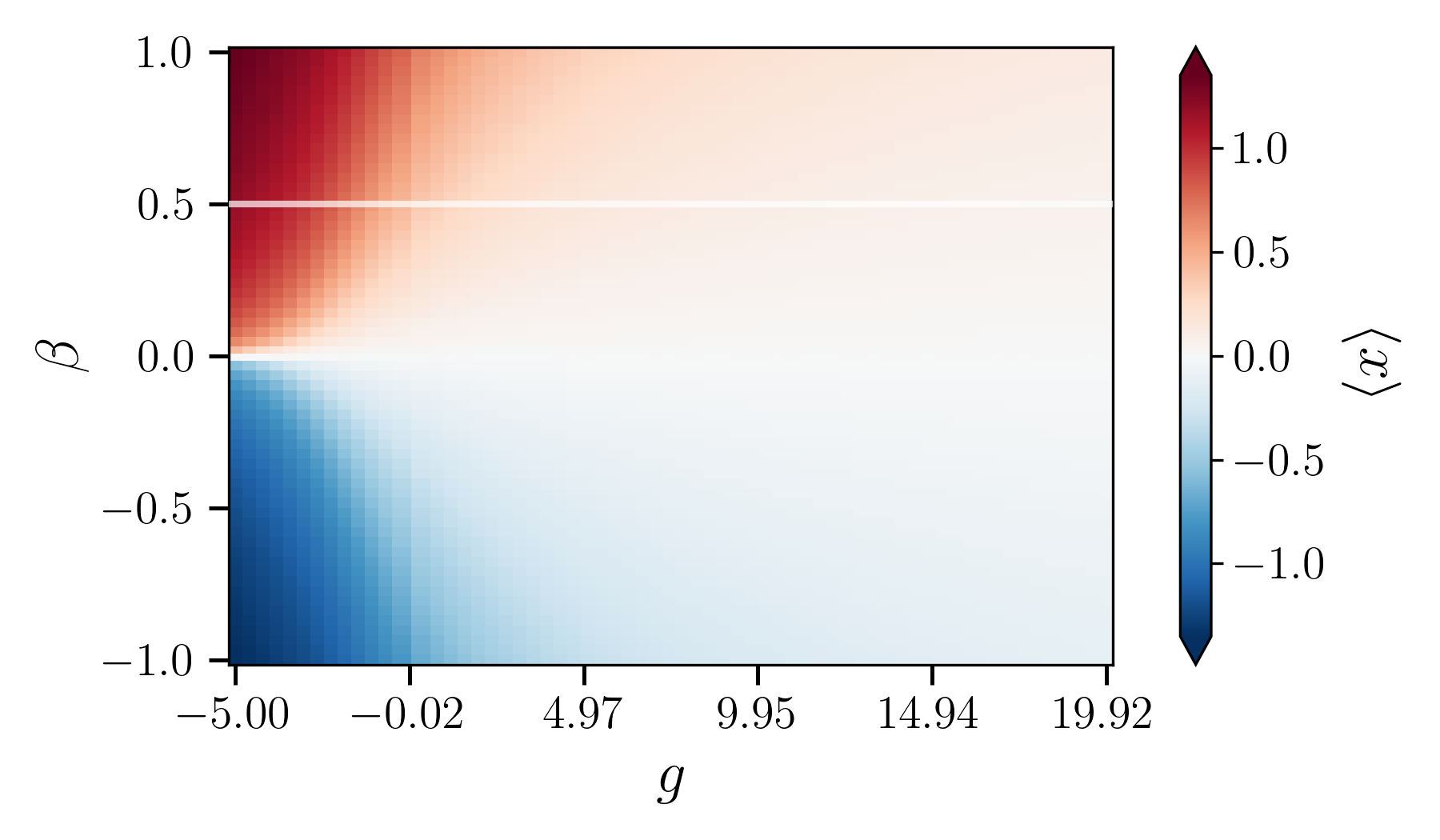}
        \put(-3,80){(b)}
        \end{overpic}
        \phantomcaption
        \label{fig:nonlinear_combined_b}
    \end{subfigure}\\[-0.4cm]
    \begin{subfigure}[b]{\linewidth}
        \centering
        \begin{overpic}[,width=\linewidth]{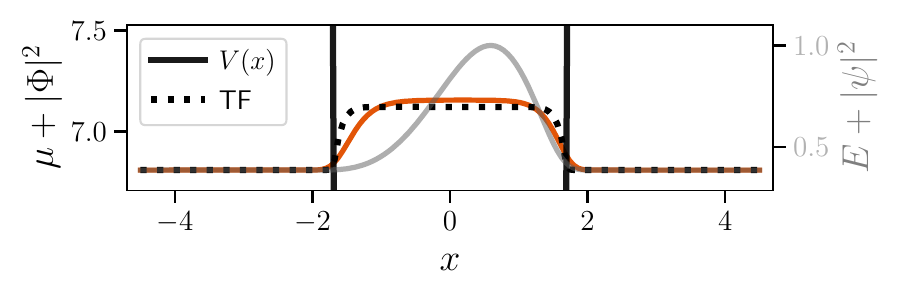}
        \put(-3,55){(c)}
        \end{overpic}
        \phantomcaption
        \label{fig:nonlinear_combined_c}
    \end{subfigure}
    \end{minipage}
    \begin{minipage}[b]{0.28\linewidth}
    \begin{subfigure}[b]{\linewidth}
        \centering
        \begin{overpic}[width=\linewidth]{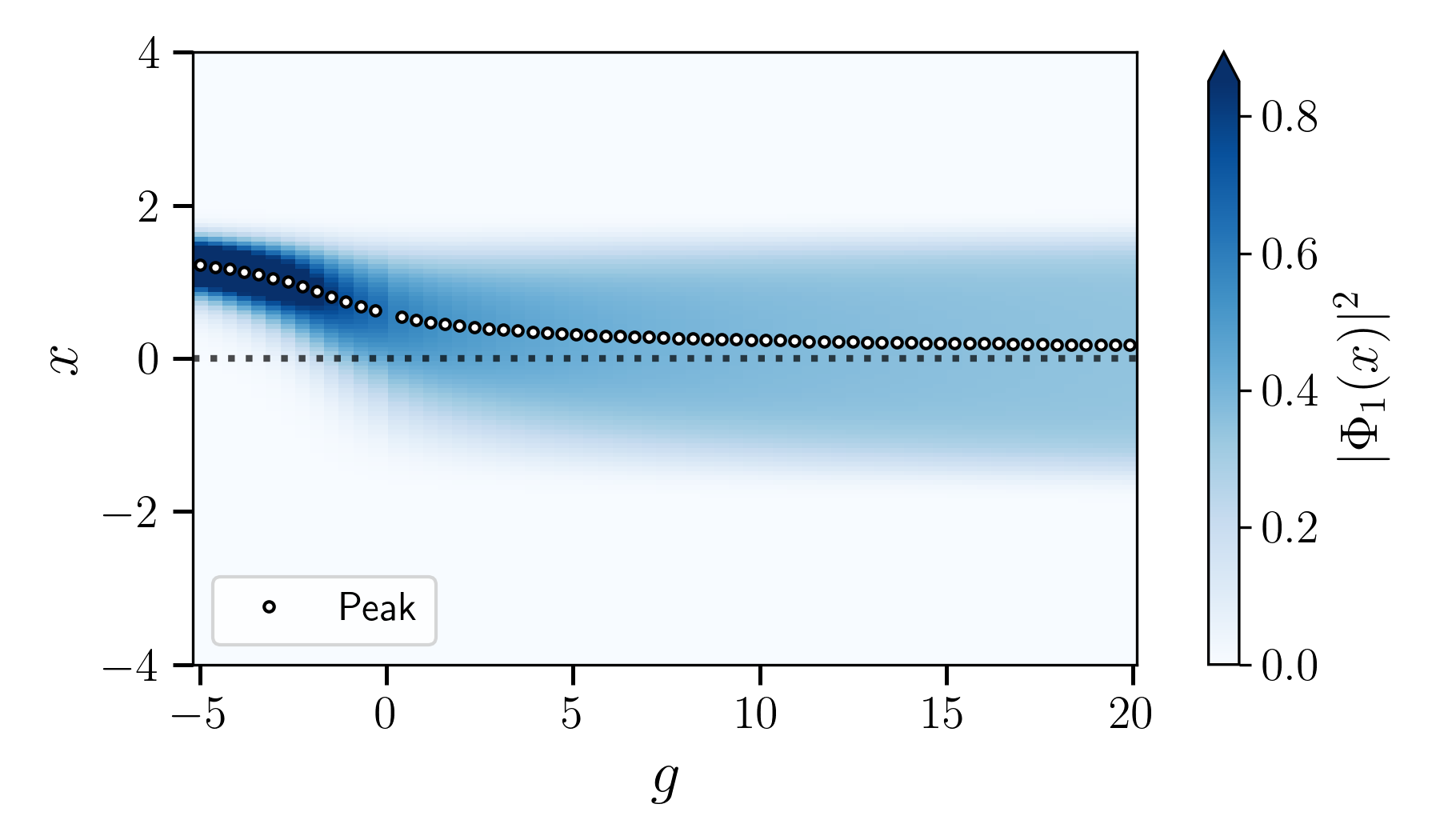}
        \put(-3,72){(d)}
        \end{overpic}
        \phantomcaption
        \label{fig:nonlinear_combined_d}
    \end{subfigure}\\[-0.9cm]
    \begin{subfigure}[b]{\linewidth}
        \centering
        \begin{overpic}[width=\linewidth]{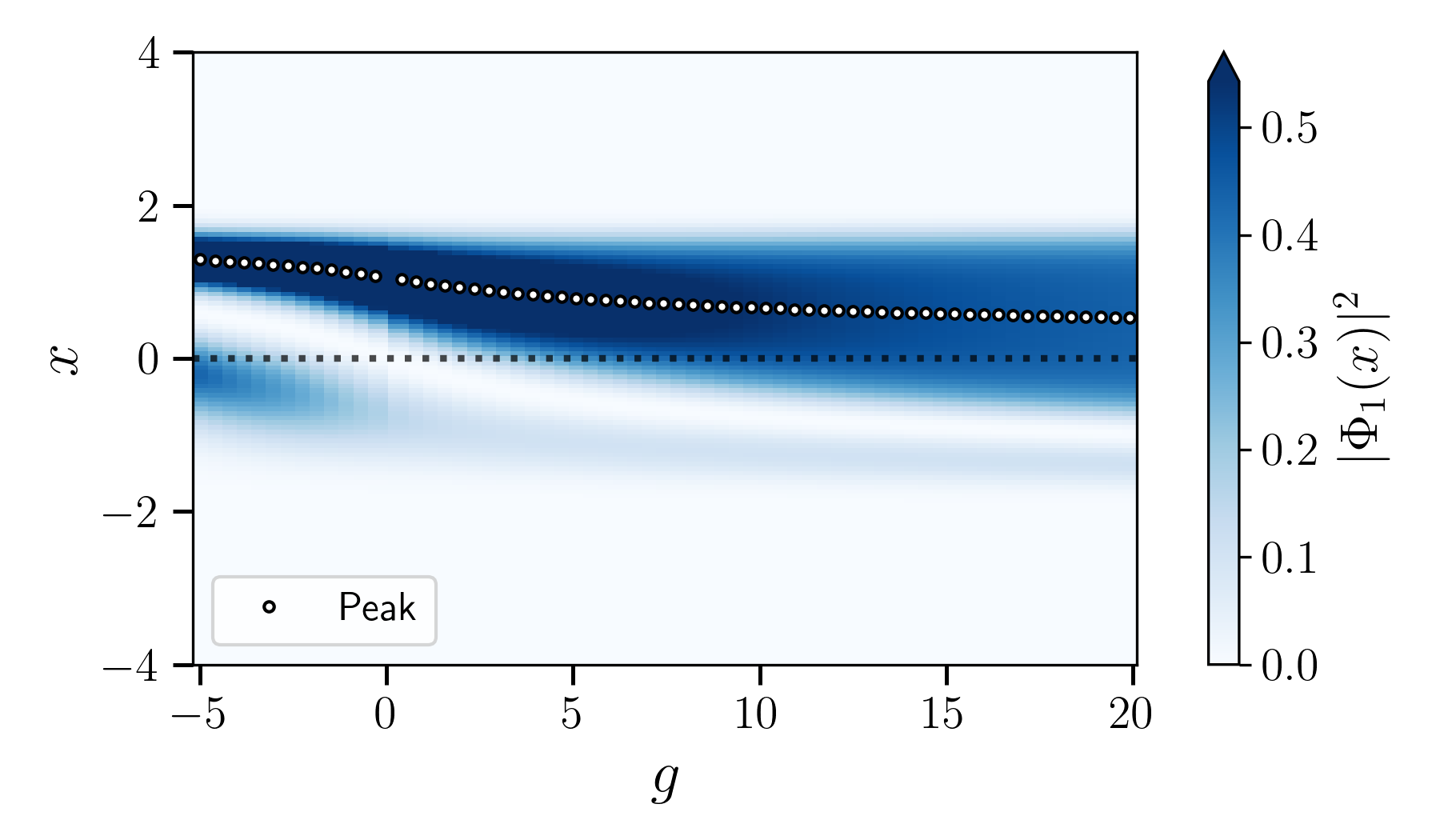}
        \put(-3, 72){(e)}
        \end{overpic}
        \phantomcaption
        \label{fig:nonlinear_combined_e}
    \end{subfigure}
    \end{minipage}\\[-0.4cm]
    \caption{ (a) First three solutions of the nonlinear eigenvalue problem for positive and negative $g=\pm 1$, in orange and purple gradients respectively, and $\beta = 0.5$. The grey curve shows the confining supergaussian trapping potential $V_{\mathrm{trap}}(x) = V_0 (1 - \mathrm{exp}\{-(x/\sigma)^m\})$ with $V_0 = 100$, $\sigma=2$ and $m=16$. 
    (b) Position expectation value $\langle x \rangle$ of the nonlinear ground state as a function of the imaginary gauge field strength $\beta$ and the interaction coupling $g$. The grey hatched region indicates the dynamically unstable regime where the BdG spectrum acquires complex eigenvalues. (c) Comparison between the ground state mode and the solution in the Thomas-Fermi limit for $g=20$ and $\beta$ as before.
    (d) (e) Ground state and first excited state amplitudes against the varying $g$ for a fixed $\beta$ indicated by the white line in panel b. The dotted black line indicates the zero for orientation, while the white dots represent the interpolated maximum value.
    }
    \label{fig:nonlinear_combined} 
\end{figure} 

\subsection{Dynamical instability}

\begin{figure}[H]
    \centering
    \includegraphics[width=0.23\linewidth]{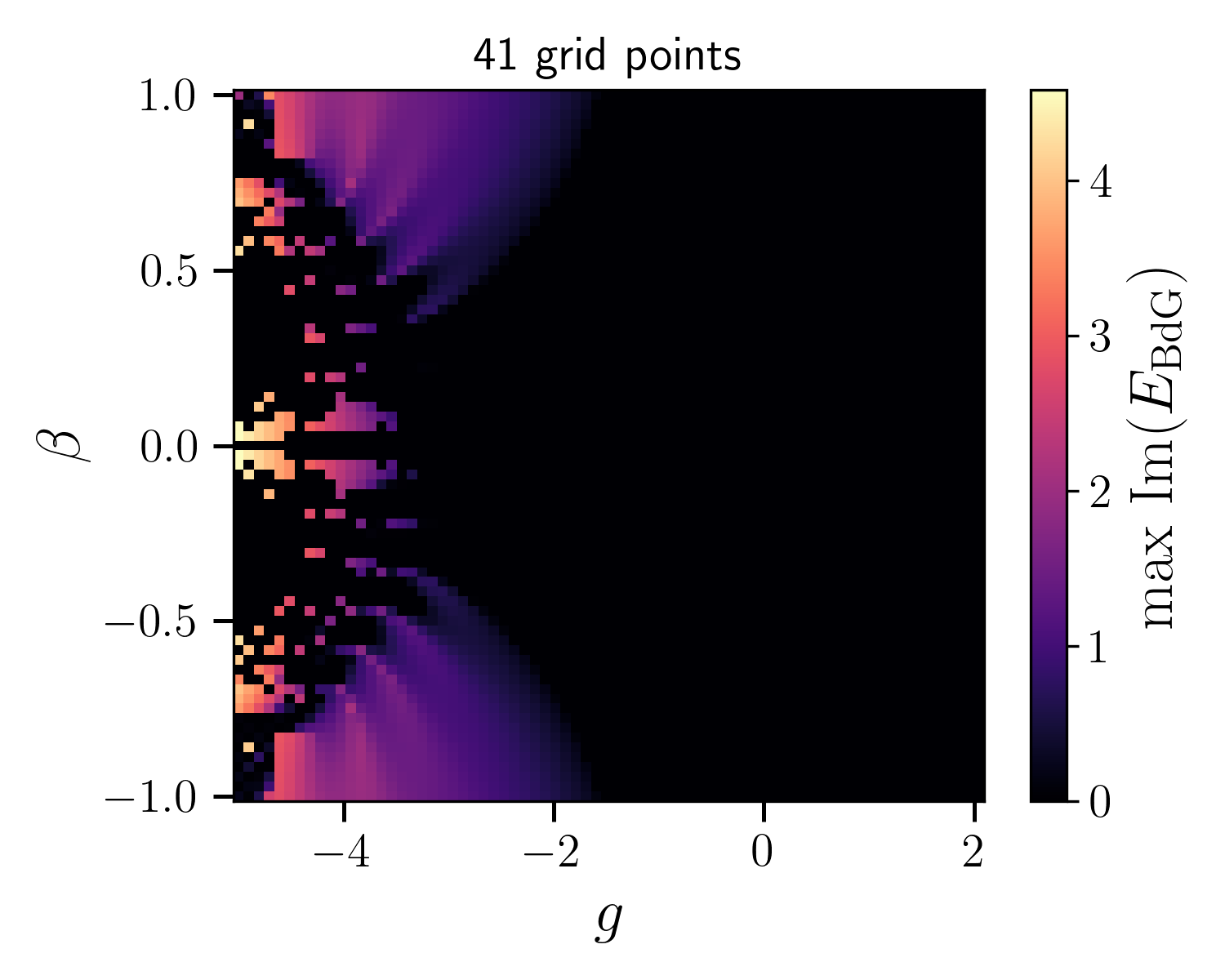}
    \includegraphics[width=0.23\linewidth]{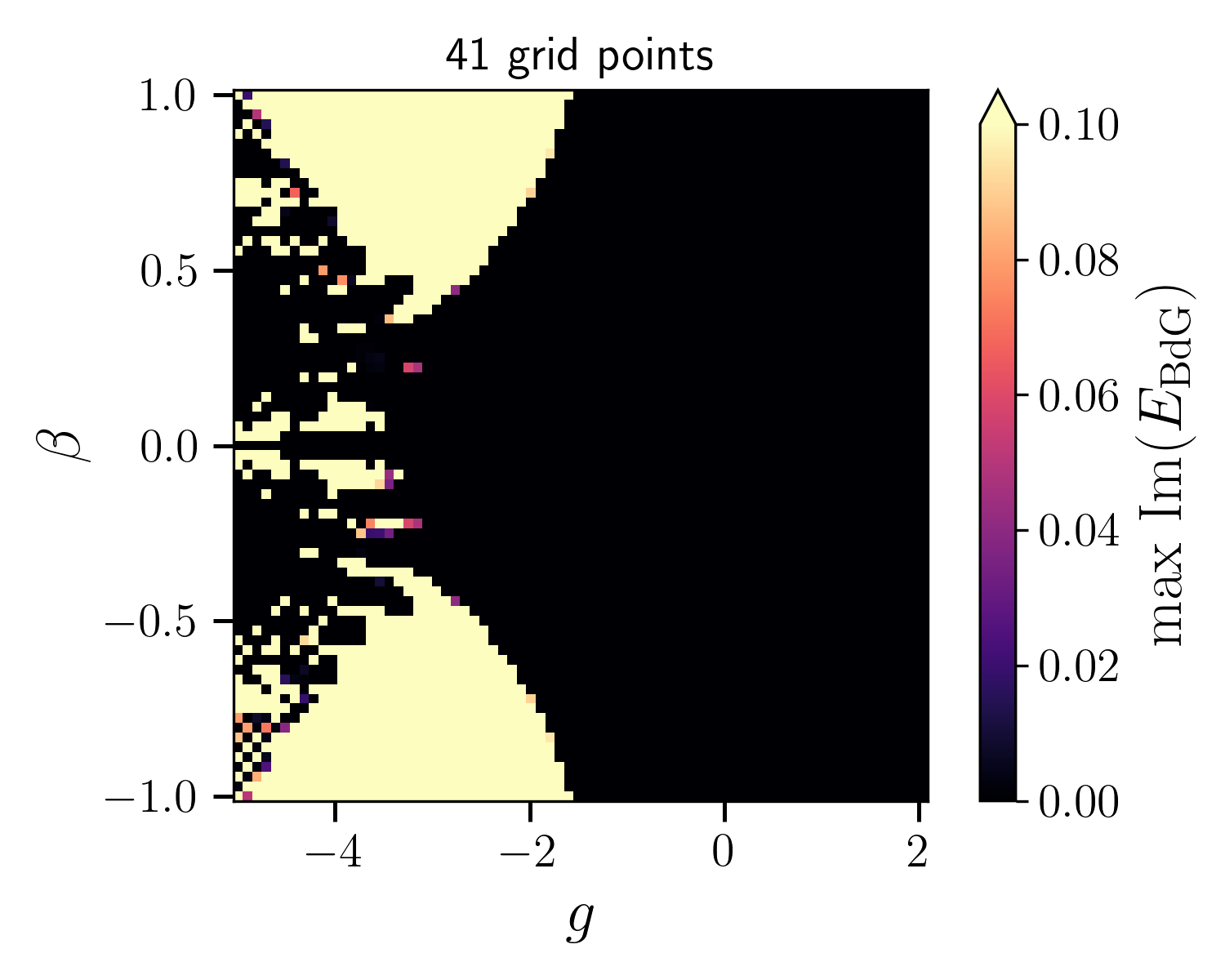} 
    \includegraphics[width=0.23\linewidth] {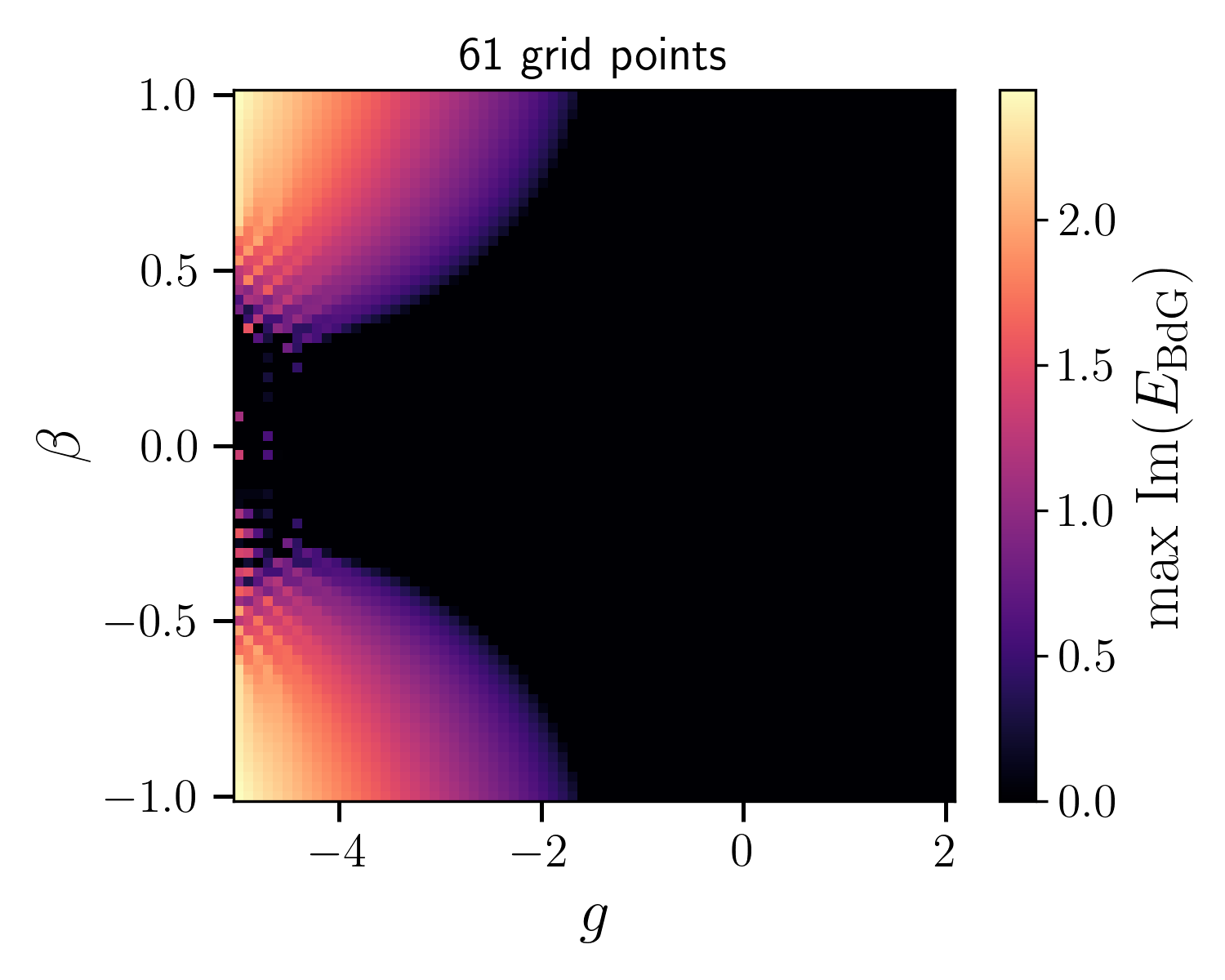}
    \includegraphics[width=0.23\linewidth]{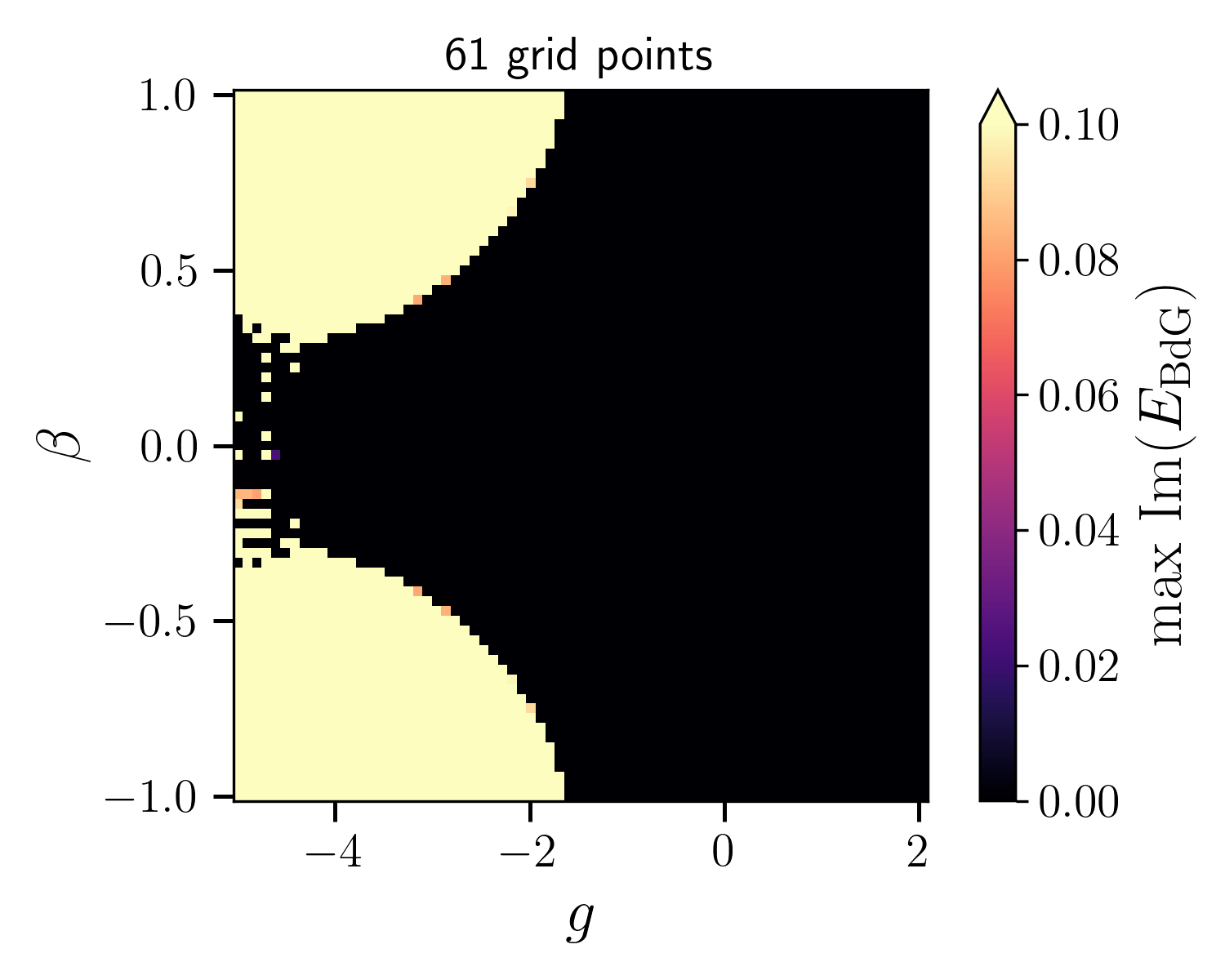} \\
    \includegraphics[width=0.23\linewidth]{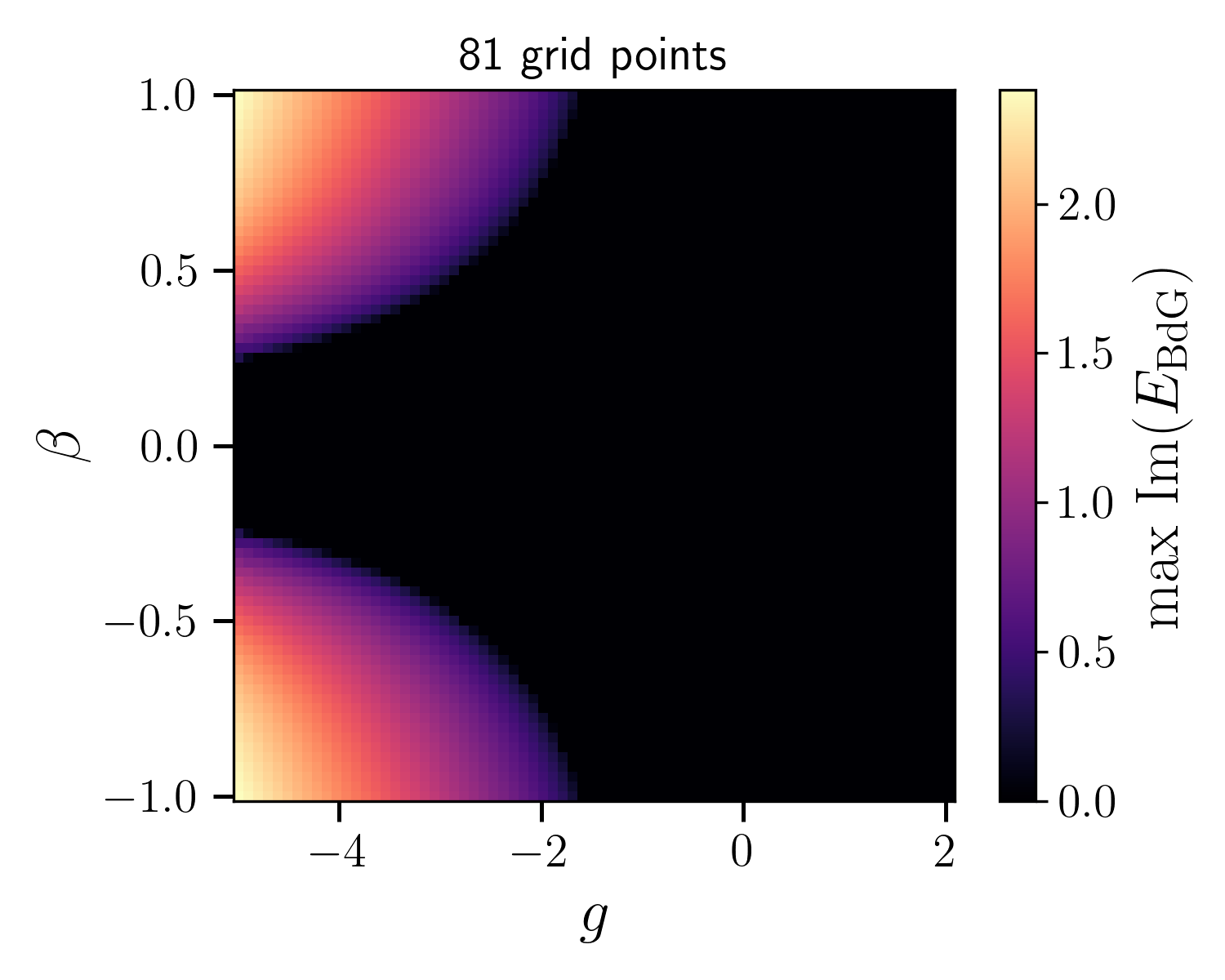}
    \includegraphics[width=0.23\linewidth]{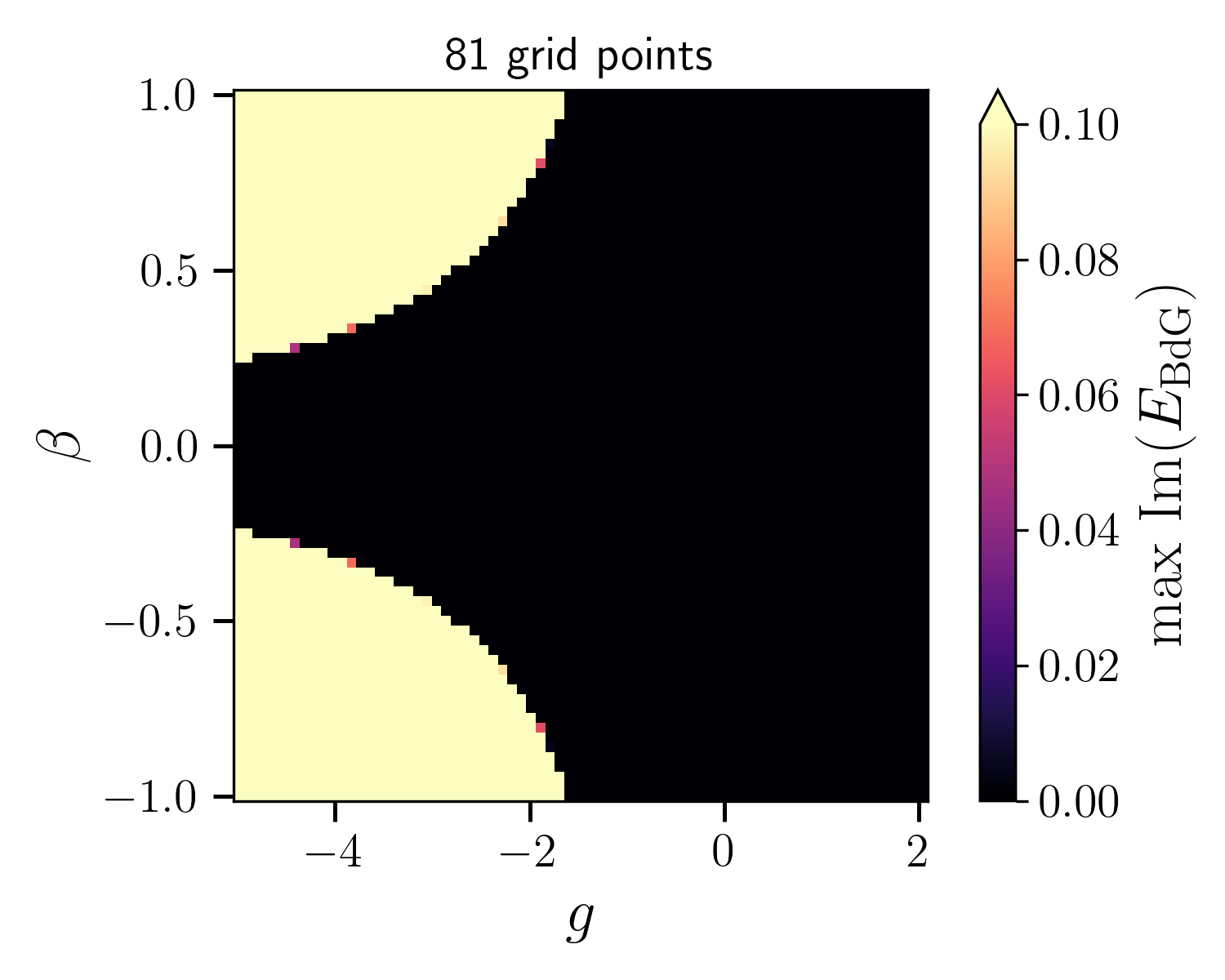} 
    \includegraphics[width=0.23\linewidth]{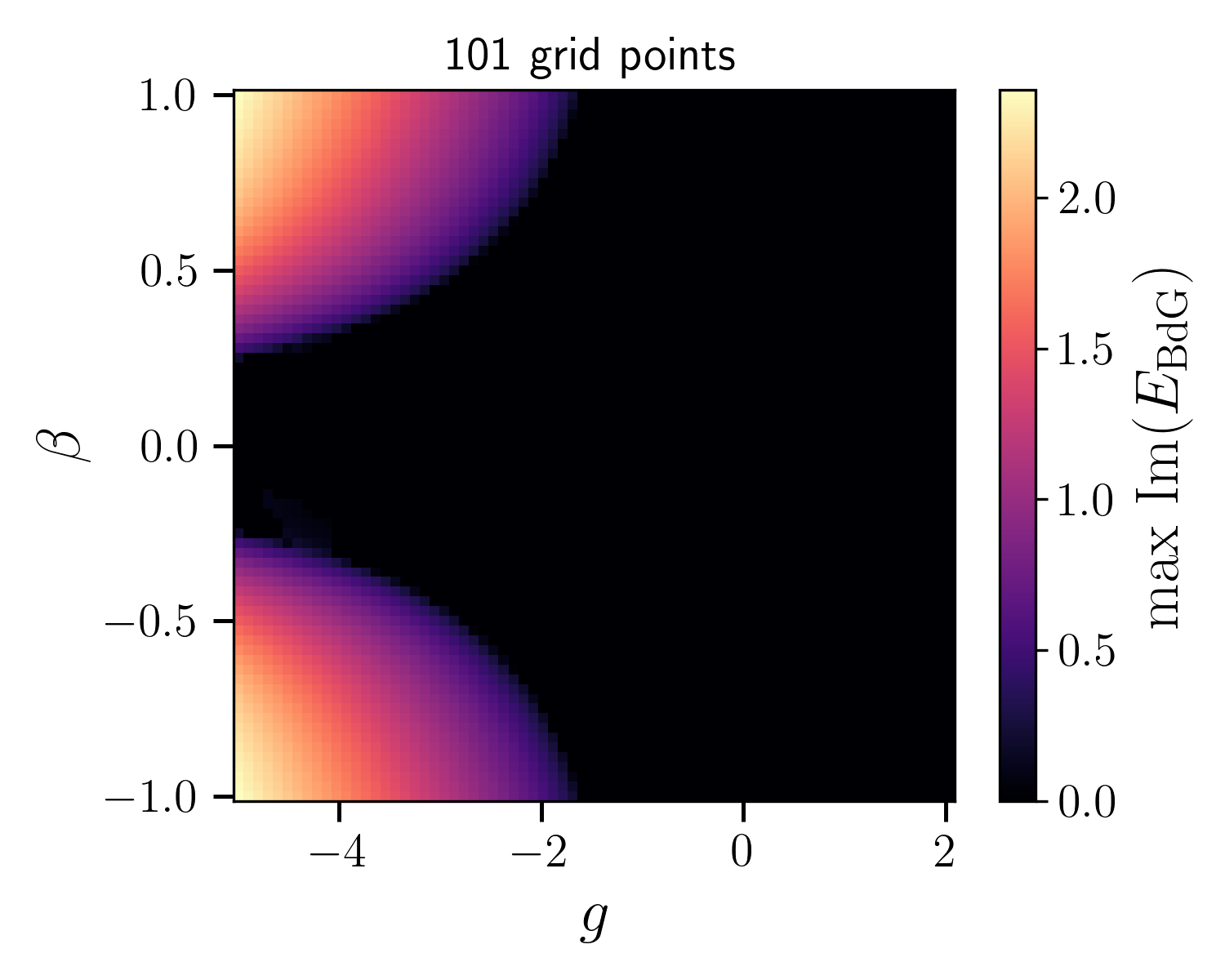}
    \includegraphics[width=0.23\linewidth]{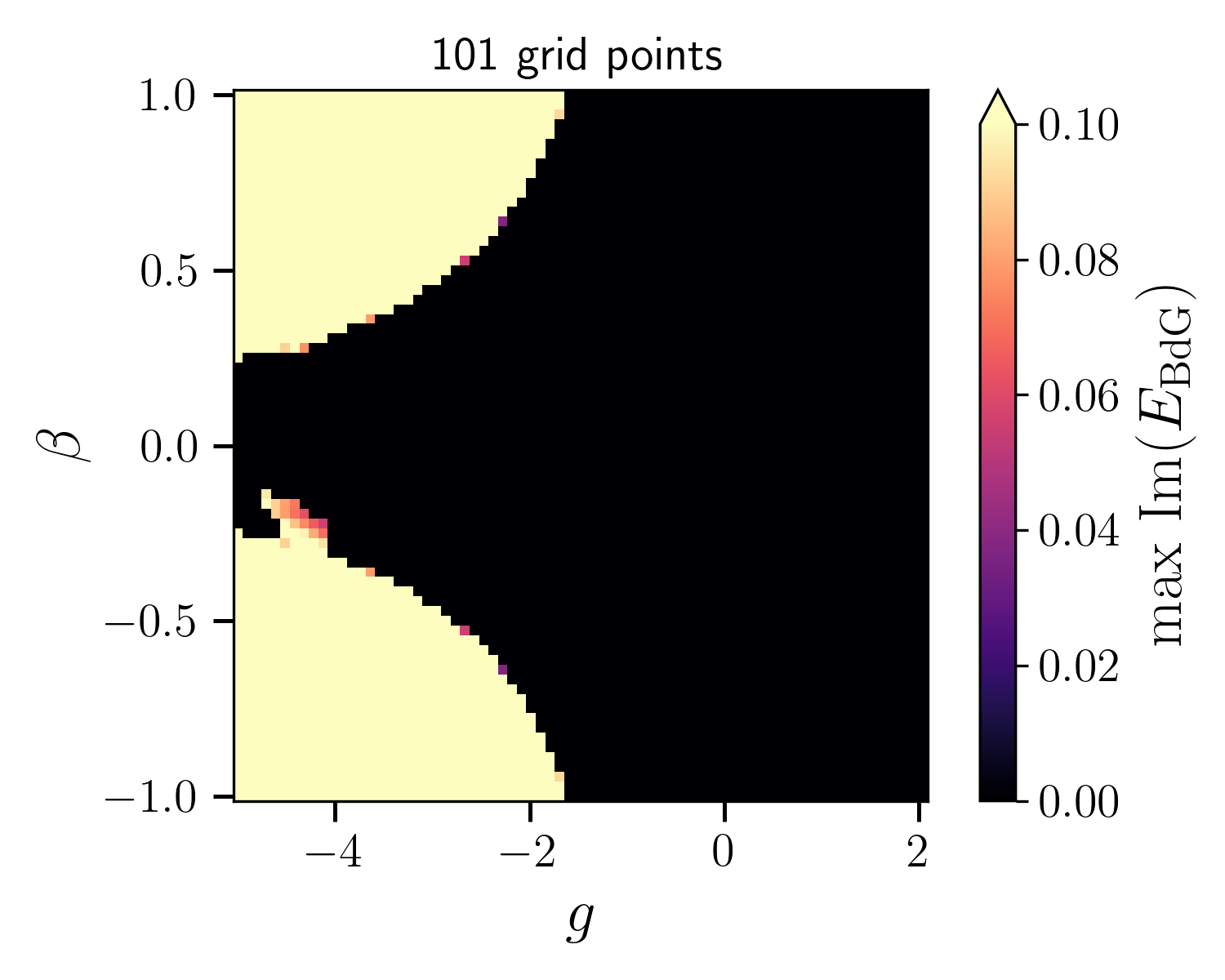} \\
    \includegraphics[width=0.23\linewidth]{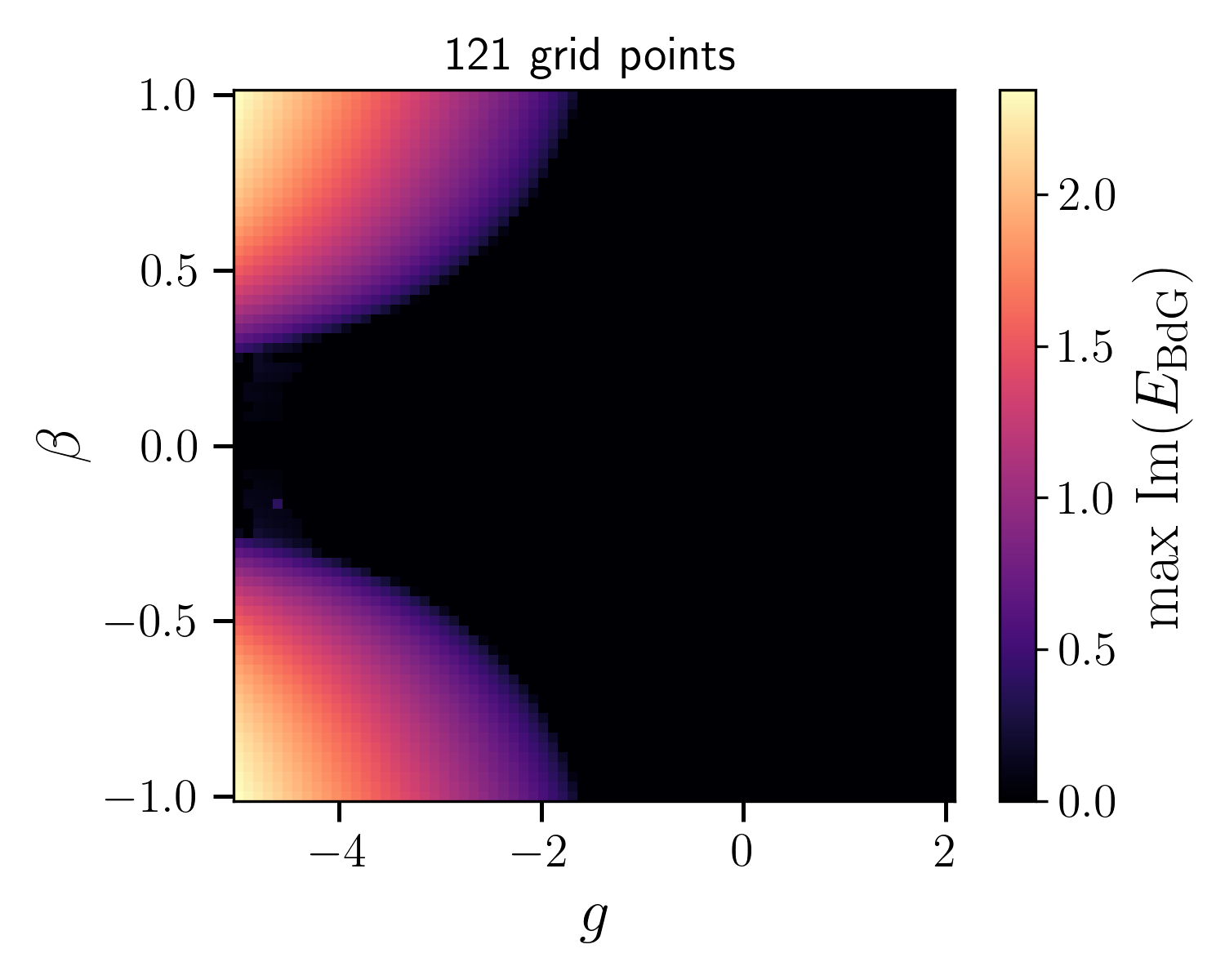}
    \includegraphics[width=0.23\linewidth]{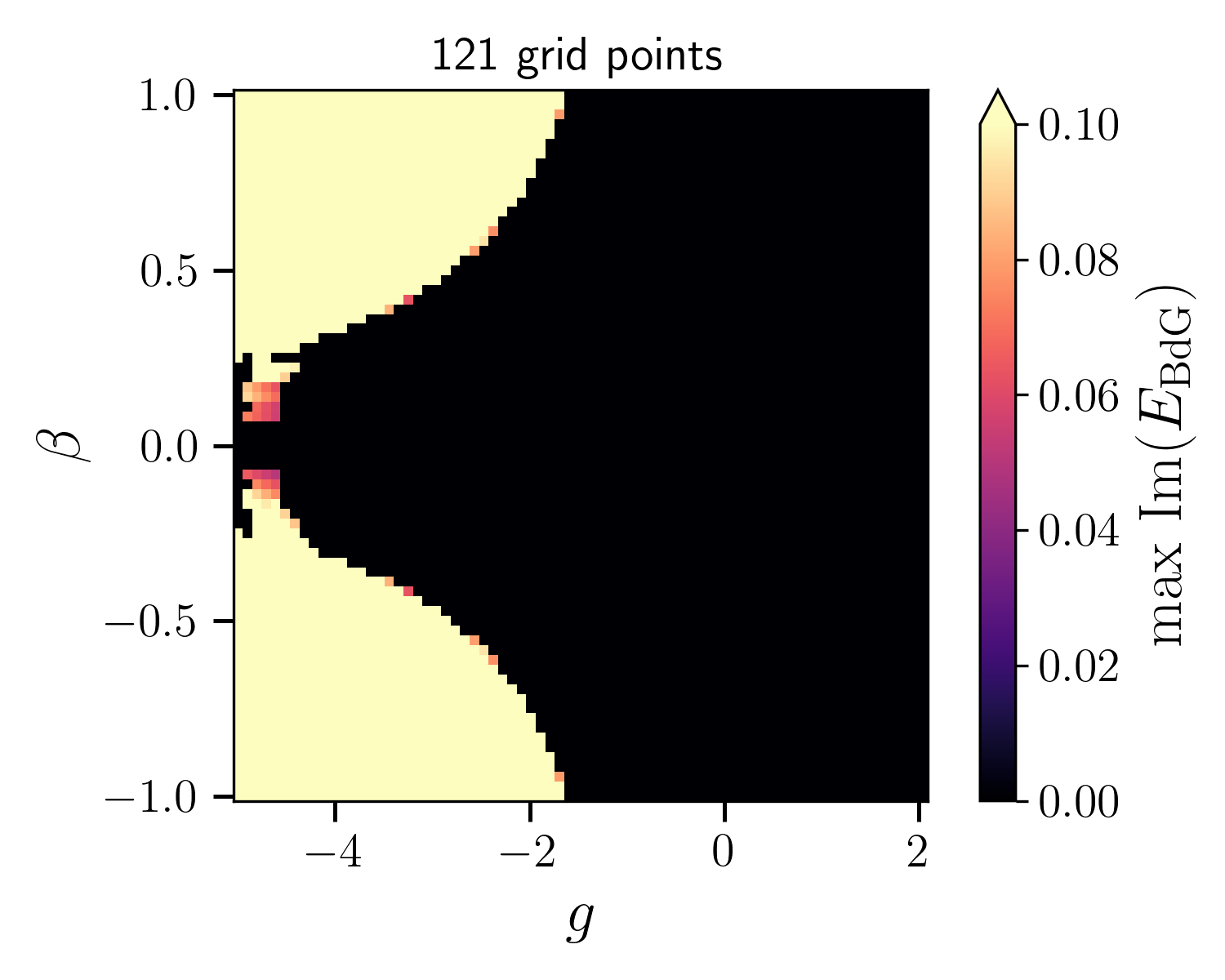} 
    \includegraphics[width=0.23\linewidth]{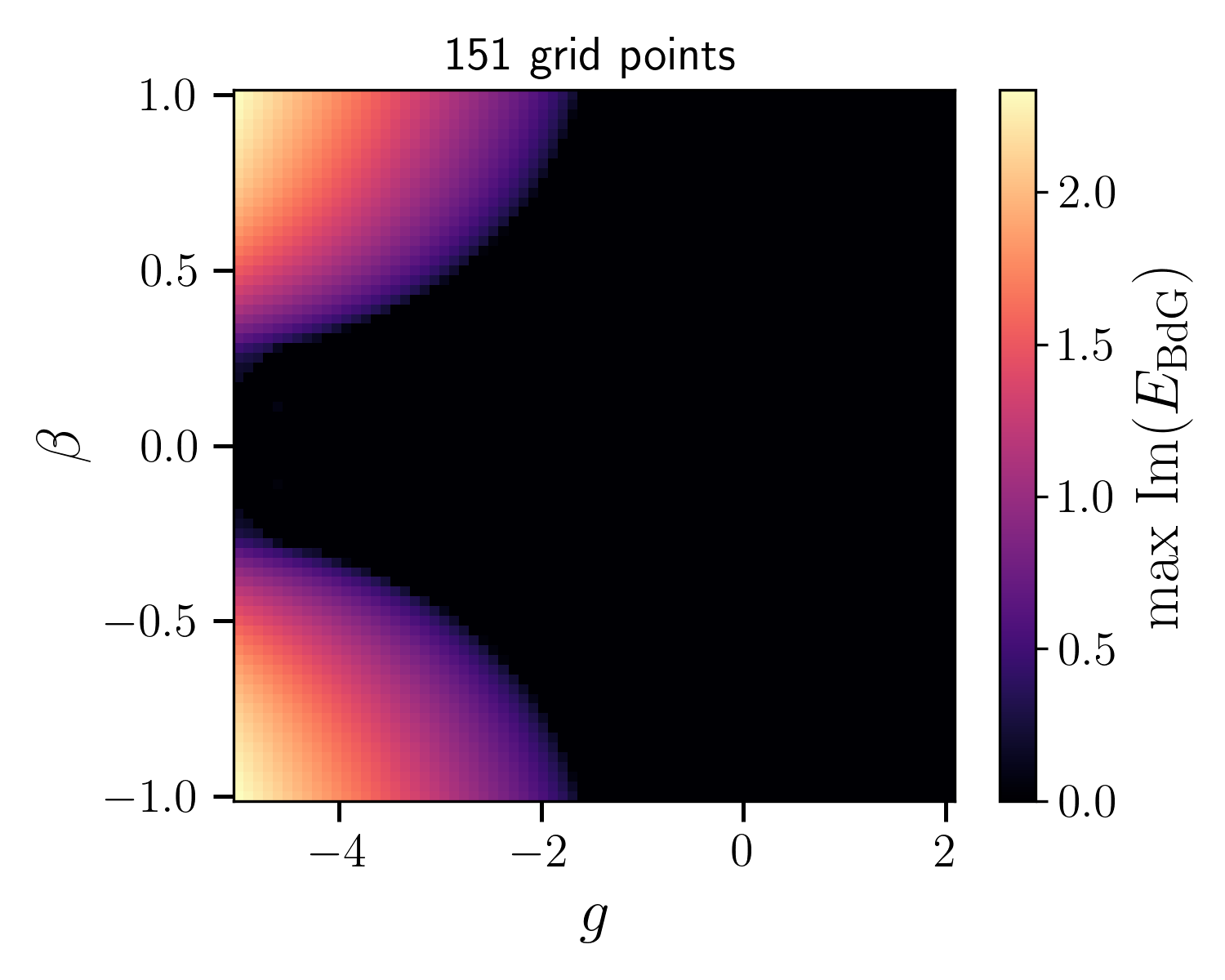}
    \includegraphics[width=0.23\linewidth]{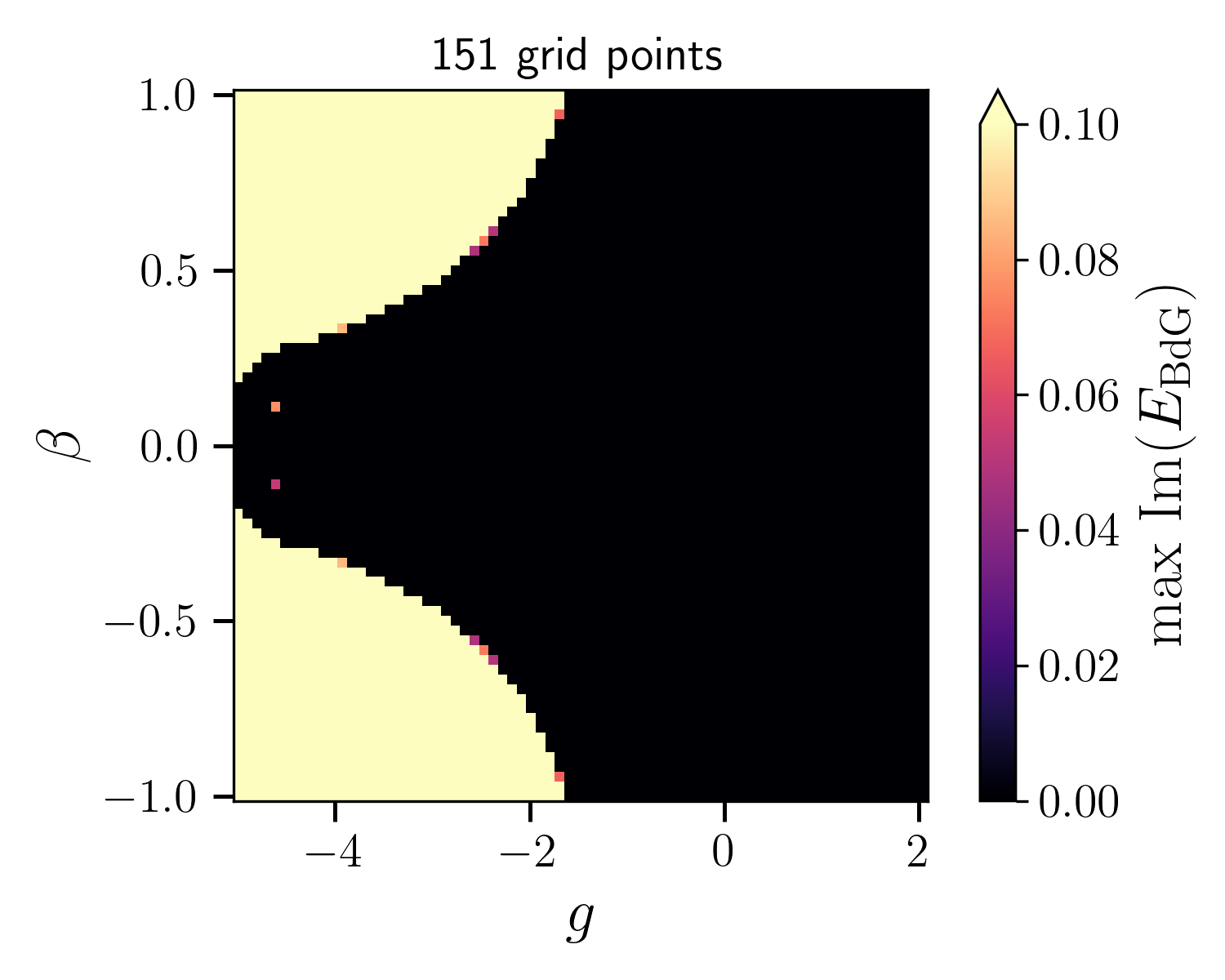}
    \caption{We vary the number of grid points and check the imaginary part of the BdG excitations in the presence of a harmonic trapping potential, and we investigate the phase space of $g$ and $\beta$.}
    \label{fig:grid_points_and_instability}
\end{figure}

\end{document}